\documentclass[twoside,11pt]{article}

\usepackage{blindtext}
\usepackage{amsthm}
\usepackage{amssymb}
\usepackage{amsmath}

\usepackage[sort,numbers]{natbib}
\usepackage[english]{babel}
\usepackage[nottoc]{tocbibind}
\usepackage[preprint]{jmlr2e}

\usepackage{jmlr2e}
\usepackage{diagbox}
\usepackage{float}
\usepackage{mathtools} 
\usepackage{multicol}
\allowdisplaybreaks[4]
\usepackage{xcolor}

\usepackage{amsfonts}
\usepackage{algorithmic}
\usepackage{graphicx}
\usepackage{textcomp}
\usepackage{dsfont}
\usepackage{booktabs}
\usepackage{hyperref}
\usepackage{subcaption}
\usepackage{setspace}

\newcommand{\R}{\mathbb{R}}
\newcommand{\Z}{\mathbb{Z}}
\newcommand{\N}{\mathbb{N}}
\newcommand{\mE}{\mathcal{E}}
\newcommand{\mL}{\mathcal{L}}

\newcommand{\mO}{\mathcal{O}}
\newcommand{\mA}{\mathcal{A}}

\newcommand{\Ta}{\Tilde{a}}
\newcommand{\Tx}{\Tilde{x}}
\newcommand{\hP}{\hat{P}}
\newcommand{\bP}{\Bar{P}}
\newcommand{\hmu}{\hat{\mu}}
\newcommand{\bmu}{\Bar{\mu}}
\newcommand{\hX}{\hat{X}}
\newcommand{\bX}{\Bar{X}}
\newcommand{\red}{\textcolor{red}}

\newcommand{\mc}{\mathcal}
\newcommand{\bR}{\mathbb{R}}
\newcommand{\bN}{\mathbb{N}}
\newcommand{\bE}{\mathbb{E}}

\newcommand{\cA}{\mathcal{A}} 
\newcommand{\cAp}{\mathcal{A}^N} 
\newcommand{\Ap}{A^N} 
\newcommand{\Kp}{K^N} 
\newcommand{\cX}{\mathcal{X}} 

\newcommand{\cN}{\mathcal{N}} 
\newcommand{\cO}{\Tilde{\mathcal{O}}}

\newcommand{\tm}[1]{t_{\operatorname{mix}}^{#1}}

\DeclareMathOperator*{\argmax}{arg\,max}

\newcommand{\eqdef}{:=}

\newcommand{\ie}{\textit{i.e.,} }

\usepackage{mathtools}
\usepackage{collect}

\definecollection{mytable}
\newcounter{duptable}
\newcounter{tmp}

\makeatletter
\newcommand\resttable{%
\let\oldlabel\label
\let\label\@gobble
\setcounter{tmp}{\value{table}}
\setcounter{table}{\value{duptable}}
\includecollection{mytable}
\let\label\oldlabel
\setcounter{table}{\value{tmp}}
}

\def\showanswers{0}

\newcommand{\hide}[1]{
\ifnum\showanswers=1
#1 \vspace{\baselineskip}
\fi
\ifnum\showanswers=0
\fi
}

\usepackage{lastpage}
\jmlrheading{}{2025}{1-\pageref{LastPage}}{}{}{}{}

\author{\name Anna Maddux\thanks{The first two authors contributed equally to this work. Correspondence to: anna.maddux@epfl.ch.} \email anna.maddux@epfl.ch \\
       \addr SYCAMORE, EPFL
       \AND
       \name $\text{Reda Ouhamma}^{\red{*}}$ \email reda.ouhamma@epfl.ch \\
       \addr SYCAMORE, EPFL
        \AND
        Hana Catic \email hana.catic@epfl.ch \\
       \addr SYCAMORE, EPFL 
        \AND
       \name Maryam Kamgarpour \email maryam.kamgarpour@epfl.ch \\
       \addr SYCAMORE, EPFL      
       }

\firstpageno{1}

\begin{document}

\title{Finite-time convergence to an $\epsilon$-efficient Nash equilibrium in potential games}

\maketitle

\begin{abstract}
This paper investigates the convergence time of log-linear learning to an $\epsilon$-efficient Nash equilibrium in potential games, where an efficient Nash equilibrium is defined as the maximizer of the potential function. Previous literature provides asymptotic convergence rates to efficient Nash equilibria, and existing finite-time rates are limited to potential games with further assumptions such as the interchangeability of players. We prove the first finite-time convergence to an $\epsilon$-efficient Nash equilibrium in general potential games. Our bounds depend polynomially on $1/\epsilon$, an improvement over previous bounds for subclasses of potential games that are exponential in $1/\epsilon$. We then strengthen our convergence result in two directions: first, we show that a variant of log-linear learning requiring a constant factor less feedback on the utility per round enjoys a similar convergence time; second, we demonstrate the robustness of our convergence guarantee if log-linear learning is subject to small perturbations such as alterations in the learning rule or noise-corrupted utilities.
\end{abstract}


\section{Introduction}

Interactions of multiple agents are at the heart of many applications, including transportation networks, auctions, telecommunication networks, and multi-robot systems. A common solution concept to describe the outcomes of multi-agent systems is the Nash equilibrium~\cite{nash1951non}. 

Thus, a natural question is whether strategic players can learn a Nash equilibrium and, if so, at what speed they can learn it.
As games can have multiple Nash equilibria of different quality, in terms of social welfare, for example, it is important to understand which Nash equilibrium is learned.


A class of games that are suitable for learning are potential games \cite{monderer1996potential}, where joint actions that maximize the potential function correspond to Nash equilibria. 
If social welfare and the potential function are aligned in the sense that an increase in social welfare is associated with an increase in potential \cite{paccagnan2022utility}, then a Nash equilibrium that maximizes the potential also maximizes the social welfare. This is the case for identical interest games and distributed welfare games \cite{marden2013distributed}, where social welfare is given by the aggregated players' utilities. 

In this paper, motivated by the connection between the potential function and social welfare, we study the speed of convergence of a decentralized learning algorithm to an approximately efficient Nash equilibrium. This question is important, as it determines how quickly desirable outcomes can be achieved through decentralized learning.



\subsection{Related work}

In potential games, many learning rules were shown to converge to an arbitrary Nash equilibrium, such as iterative best-response dynamics \cite{rosenthal1973class,awerbuch2008fast,chien2011convergence}, no-regret algorithms \cite{krichene2015online, palaiopanos2017multiplicative,heliou2017learning}, and fictitious play \cite{monderer1996fictitious,monderer1996potential}. On the other hand, log-linear learning \cite{blume1993statistical,young1993evolution} is the only known algorithm to converge to a specific Nash equilibrium, namely, the potential maximizer. 
Past works provide asymptotic convergence guarantees \cite{blume1993statistical, young1993evolution} as well as an asymptotic rate of convergence to a potential function maximizer \cite{tatarenko2017game}. 
Such asymptotic convergence guarantees were also shown for slight variations of log-linear learning which include synchronous updates \cite{marden2012revisiting}, utilities that are corrupted by noise \cite{leslie2011equilibrium}, or payoff-based or two points of feedback per round \cite{marden2012revisiting,arslan2007autonomous}
Another line of research studied the mixing time of the Markov chain induced by log-linear learning \cite{auletta2013mixing, auletta2011convergence} and its transient behavior prior to reaching the stationary distribution \cite{auletta2012metastability}, but did not explicitly relate the stationary distribution to the set of efficient Nash equilibria.


While asymptotic convergence and convergence rates characterize the long-run behavior and the asymptotic speed, respectively,  at which log-linear learning approaches an efficient equilibrium, finite-time convergence provides explicit bounds on the number of steps required to reach an $\epsilon$-efficient Nash equilibrium, making it particularly desirable in practice.

Few past works provide finite-time guarantees for log-linear learning to an $\epsilon$-efficient Nash equilibrium, an action profile whose potential is $\epsilon$-close to the maximum value. In specific potential games, namely, in atomic routing games with polynomial costs, \cite{asadpour2009inefficiency} derives a convergence time that is exponential in $1/\epsilon$ and polynomial in $N$, the number of players. Moreover, in games with a graph structure between players, \cite{montanari2008convergence,montanari2010spread} prove a convergence time, which is exponential in $1/\epsilon$ and $N$ in the worst case. Finally, in potential games with interchangeable players and a Lipschitz-continuous potential function, \cite{shah2010dynamics} shows a convergence time exponential in $A$ and $1/\epsilon$ and linear in $N$, where $A$ is the number of actions per player. The latter result was extended to semi-anonymous potential games  \cite{borowski2015fast}, which consist of groups of interchangeable players.

 While finite-time convergence of log-linear learning has been proven for subclasses of potential games, to our knowledge, it is not established for general potential games.

\subsection{Contributions}


In this paper, we derive the first finite-time convergence guarantees of log-linear learning to an $\epsilon$-efficient Nash equilibrium in general potential games. 
Our contributions are:
\begin{itemize}
    \item We prove a convergence time of 
    $
    \cO((A^N / \epsilon)^{\frac{1}{\Delta}})
    $ to an $\epsilon$-efficient Nash equilibrium (Theorem \ref{thm:convergence_rates}), where the suboptimality gap $\Delta$ is a problem-dependent constant. 
    \item If in addition, the players are interchangeable, then an $\epsilon$-Nash equilibrium is reached in $\cO((\frac{N^A}{\epsilon})^{\frac{1}{\Delta}})$ which in contrast to general potential games is polynomial in $N$ (Corollary \ref{cor:symmetric_potential}). 
    \item We consider two variants of log-linear learning: binary log-linear learning and perturbed log-linear learning motivated by limited feedback and noise corrupted utilities, respectively. We prove a convergence time of $
    \cO((A^N / \epsilon)^{\frac{1}{\Delta}})
    $ (Theorem \ref{thm:convergence_rates_binary}) and $\cO((A^N / \epsilon)^{\frac{1}{\Delta}N(1+\xi)})$ (Theorem \ref{thm:General_Convergence}) , respectively.
\end{itemize}

On the technical side, past works \cite{blume1993statistical, young1993evolution} established that log-linear learning induces a Markov chain. To obtain our novel finite-time results, we build on this connection and develop new results about Markov chain mixing times and stationary distributions as follows:
\begin{itemize}
    \item We use mixing-time bounds based on the so-called log-Sobolev constant of the Markov chain \cite{diaconis1996logarithmic} to establish finite-time convergence guarantees for log-linear learning and its variants. To this end, we derive a novel bound on the log-Sobolev constant of a class of Markov chains which includes those induced by log-linear- and binary log-linear learning (Lemma \ref{lem:general_bound_log_Sobolev}).
    \item We derive a tight Lipschitz constant of stationary distributions of Markov chains as a function of their transition matrix (Lemma \ref{lem:Lipschitz}). 
    We leverage this result to study the convergence of perturbed log-linear learning for which the stationary distribution is unknown (Theorem \ref{thm:General_Convergence}).
\end{itemize}

\noindent
\textit{Notations:} We denote by $[N]$ the set $\{1, \ldots, N\}$. For a finite set $\cX$, we denote by $\Delta(\cX)$ the probability simplex over $\cX$, and by $\mathds{1}_{a \in \cX}$ the indicator function of $\cX$. Finally, we use the big-$\mO$ notations $\cO$ and $\Tilde{\Omega}$ to hide logarithmic terms.

\section{Problem setup}




We consider a potential game with $N$ players. Every player has an action set $\mA$ of cardinality $A <\infty$, which for simplicity we assume to be the same for all players. The utility of player $i$ is a mapping $U_i:\cAp\rightarrow[0,1]$, where $\cAp$ is the joint action space. In a potential game, the utility functions are characterized by a potential function $\Phi:\cAp \rightarrow \bR$ such that:

    \begin{align*}
    U_i(a_i,a_{-i})-U_i(a_i',a_{-i})=\Phi(a_i,a_{-i})-\Phi(a_i',a_{-i}),\quad \forall i\in[N], \forall a_i,a_i'\in\cA, \forall a_{-i}\in \cA^{N-1}.
\end{align*}


A common solution concept is the Nash equilibrium \cite{nash1950equilibrium}, at which no player can improve her utility by unilaterally changing her action.

\begin{definition}

    A Nash equilibrium is an action profile $(a_i)_{i\in [N]} \in \cAp$ that satisfies:
    
    \begin{align*}
        U_i(\bar{a}_i,a_{-i}) \le U_i(a_i,a_{-i}),\quad \forall i \in [N], \forall \bar{a}_i\in \cA,
    \end{align*}
    where $a_{-i}\eqdef(a_j)_{j \in [N]\setminus \{i\}}$ is the action of all players but $i$.
\end{definition}

Generally, a game may have several Nash equilibria, as shown in the two-player potential game below. Here, action profiles $(A, A)$ and $(B, B)$ are both Nash equilibria. However, the value of the potential function may differ for different Nash equilibria, as is the case in the example.

\begin{align*}
      \bordermatrix{ & A & B &  \cr
        A & (5,2) & (-1,-2) \!\!\!\!\!  \cr
        B & (-5,-4) & (1,4) \!\!\!\!\!\!},
        \quad \text{with potential:} \quad
        \bordermatrix{ & A & B &  \cr
        A & 4 & 0 \!\!\!\!\!\!\!\! \cr
        B & -6 & 2 \!\!\!\!\!\!\!\! }. 
\end{align*}
This example motivates the distinction between a Nash equilibrium and an \emph{efficient} Nash equilibrium, defined as:

\begin{align}\label{def:efficient_NE}
    a^* \in \argmax_{a\in\cAp}\Phi(a).
\end{align}
Note that such $a^*$ exists and is a Nash equilibrium~\cite{monderer1996potential}. 

In a game setting,  players act independently and do not have knowledge of the other players' utilities, nor do they share their utility with a central authority. As a result, it is impossible to enumerate over the set of joint actions to identify the potential function and thus find its maximizer. 

Instead, in this work, we consider a repeated game setting, where the game unfolds over multiple rounds. Our focus is on learning rules that converge to an $\epsilon$-efficient Nash equilibrium.
\begin{definition}
    An action profile $a^*\in\mc A^N$ is an $\epsilon$-efficient Nash equilibrium if it satisfies $\Phi(a^*) \ge \max_{a \in \mathcal{A}^N} \Phi(a) - \epsilon$ for $\epsilon\in(0,1)$.
\end{definition} 

\section{Convergence of log-linear learning}

\noindent
In this section, we introduce the well-established log-linear learning rule \cite{blume1993statistical} and state our main result on the convergence time of log-linear learning to an $\epsilon$-efficient Nash equilibrium. 

\subsection{Algorithm and background}

We consider a repeated game setting in which all players follow log-linear learning. In the initial round, players initialize their action randomly according to some distribution $\mu^0$. Thereafter, at round $t$, a player denoted by $i$ is randomly chosen among all players and allowed to alter her action while the other players repeat their current action, \ie $a_{-i}^t=a_{-i}^{t-1}$. Player $i$ observes her utility for all actions $a_i\in\cA$ given the other players' actions $a_{-i}^{t-1}$.  
Then, player $i$ samples an action from her strategy $p_i^t\in\Delta(\cA)$ such that:

\begin{align}\label{eq:log_linear_update}
    p_i^t(a_i)=\frac{e^{\beta U_i(a_i,a_{-i}^{t-1})}}{\sum_{a_i'\in\mathcal{A}}e^{\beta U_i(a_i',a_{-i}^{t-1})}},\quad \forall a_i\in\cA,
\end{align} 
where parameter $\beta$ measures rationality: for large $\beta$ player $i$ likely selects a best response $a_i^{t}\in\argmax_{a_i\in\mA}U_i(a_i, a_{-i}^{t-1})$; and for $\beta = 0$ player $i$ samples $a_i^t$ uniformly. 

Log-linear learning induces an irreducible and aperiodic Markov chain $\{X_t\}_{t\in\Z_+}$ over state space $\cAp$ with a time-reversible transition matrix $P\in\R^{A^N\times A^N}$ \cite{marden2012revisiting} given by:
\begin{align}\label{eq:transition_prob_log_linear}
    P_{a,\Ta} = \frac{1}{N}\frac{e^{\beta U_i(\Ta_i,\Ta_{-i})}}{\sum_{a_i'\in\mathcal{A}}e^{\beta U_i(a_i',\Ta_{-i})}} \mathds{1}_{\Ta \in \cN(a)}. 
\end{align}
Here $\cN(a) = \{\Ta\in\cAp \: | \:\exists i\in [N]: \Ta_{-i} = a_{-i}\}$ is the set of action profiles $\Ta\in\cAp$ that differ from action profile $a\in\cAp$ in at most one player's action. The stationary distribution $\mu\in\Delta(\cAp)$ of log-linear learning is given by \cite{blume1993statistical}:

\begin{align}\label{eq:stationary_dist_log_linear}
    \mu(a) = \frac{e^{\beta\Phi(a)}}{\sum_{\Ta\in\cAp}e^{\beta\Phi(\Ta)}}, \quad\forall a\in \cAp. 
\end{align}

\noindent The above can be verified by checking the detailed balance equations corresponding to $\mu$ \cite{montenegro2006mathematical}.\footnote{Detailed balance holds if $\mu(a)P_{a,\Ta}=\mu(\Ta)P_{\Ta,a}$ for all $a,\Ta\in\Ap$.} It follows that we can analyze the convergence time of log-linear learning by studying the associated Markov chain. 

Previous works \cite{blume1993statistical, young1993evolution, marden2012revisiting} show that for sufficiently large $\beta$ log-linear learning converges asymptotically to a potential function maximizer and thus to an efficient Nash equilibrium. With the exception of a few works \cite{montanari2010spread,asadpour2009inefficiency,shah2010dynamics} which make additional assumptions on the potential game, none of the previous works, however, provide finite-time convergence guarantees. Thus, in the following section, we establish our main result on the convergence time of log-linear learning to an $\epsilon$-efficient Nash equilibrium in general potential games.

\subsection{Convergence time in general potential games}

We first introduce some notation needed to state our main result. Denote by $a^*$ a potential maximizer, namely $a^*\in\arg\max_{a\in\cAp}\Phi(a)$ and  by $\mc A_*^N \eqdef \{a^*\in\cAp | a^*\in\arg\max_{a\in\mc A^N}\Phi(a)\}$ the set of optimal action profiles with cardinality $A_*^N = |\mc A_*^N|$. Moreover, we define the suboptimality gap as
\[
\Delta \eqdef \min_{a \in \mathcal{A}_P : \Phi(a) < \Phi(a^*)}
\bigl(\Phi(a^*) - \Phi(a)\bigr).
\]
By construction, $\Delta \ge 0$. The degenerate case $\Delta = 0$ is trivial, since it implies that all action profiles achieve the optimal potential value and hence are efficient Nash equilibria. Consequently, throughout the remainder of this paper, we assume $\Delta > 0$.


\begin{theorem}\label{thm:convergence_rates}
    Consider a potential game with a potential function $\Phi:\cAp\rightarrow [0,1]$ and with $A\ge 4$.\footnote{We assume $A\ge 4$ to bound the log-Sobolev constant in Lemma \ref{lem:general_bound_log_Sobolev}.} For $\epsilon\in(0,1)$ and initial distribution $\mu^0$, assume that players adhere to log-linear learning with:

    \begin{align}\label{eq:beta}
        \beta\!&\geq \!\frac{1}{\Delta}\!\log\left(\!\!\left(\Ap\!\!-\!A_*^N\right)\!\!\left(\!\frac{2}{\epsilon A_*^N}\!-\!\frac{1}{A_*^N}\right)\!\right)\!.\!
    \end{align}
     Then,  

    \begin{align*}
        \mathbb{E}_{a\sim \mu^t}[\Phi(a)]\geq \max_{a\in\cAp}\Phi(a)-\epsilon,
    \end{align*}
    for $t\geq \frac{25N^2 A^5}{16\pi^2 }e^{4\beta}\bigg(\log\log \Ap+\log \beta+2\log\frac{4}{\epsilon}\bigg)$.
\end{theorem}

\noindent In other words, after {\small$t=\cO(N^2A^5(\frac{A^N}{\epsilon})^{1/\Delta})$} rounds of log-linear learning with {\small$\beta = \Tilde{\Omega}\left(\frac{1}{\Delta}\log \frac{\Ap}{\epsilon}\right)$} the expected potential function value of the joint action at time $t$ is $\epsilon$-optimal. We provide a proof of this Theorem in Section \ref{sec:proof_main_theorem}.

Here, we discuss the result. Theorem \ref{thm:convergence_rates} provides the first finite-time convergence rate to an $\epsilon$-efficient Nash equilibrium in general potential games. 
For $\epsilon\in(0,1)$, the convergence time grows polynomially in $A$ and $1/\epsilon$ and exponentially in $N$. We note that the exponential dependence on $N$ is unavoidable since the problem of finding an efficient Nash equilibrium in a potential game is NP-hard. This was shown for the integral multicast game and the fair cost-sharing game, which are instances of a potential game \cite[Theorem~5]{chekuri2006non} and \cite[Theorem~9]{balcan2013circumventing}, respectively. 
Thus, the exponential dependence on $N$ in our bounds reflects the intrinsic complexity of the problem, rather than a limitation of the chosen learning rule or the convergence analysis.
However, to our knowledge, we are the first to avoid exponential dependence on $1/\epsilon$ (see Table \ref{tab:sample-table}), which we achieve by introducing the problem-dependent constant $\Delta$. Similar suboptimality-based notions have been widely used in the stochastic multi-armed bandit literature, both for regret analysis \cite{auer2002finite} and best-arm identification \cite{karnin2013almost}. While direct computation of $\Delta$ is infeasible without access to the potential function, it can often be estimated using domain knowledge or sampling-based techniques.

\begin{table}
\setlength{\tabcolsep}{3pt}
\caption{Convergence of log-linear learning to $\epsilon$-efficient Nash equilibrium.}
  \label{tab:sample-table}
  \centering
  \begin{tabular}{lll}
    \toprule
     \textbf{Game setting} & \textbf{Assumptions}   & \textbf{Convergence time}     \\
    \toprule
     Routing game with  & Cost functions of & {\tiny$\cO(e^{\frac{N}{\epsilon}})$}  \\
     $K$ vertices \cite{asadpour2009inefficiency} & degree at most $p$ &\\
     \midrule
    Potential game with inter-  & $\lambda$-Lipschitz  & {\tiny$\cO(N (\frac{A\lambda}{\epsilon})^{\frac{A}{\epsilon}})$} \\ 
    changeable players \cite{shah2010dynamics} & continuous potential&\\
    \midrule
    Corollary \ref{cor:symmetric_potential}  & $A\ge 4$ & {\tiny$\cO(N(\frac{N^A}{\epsilon})^{\frac{1}{\Delta}})$}
    \\
    \midrule
    Theorem \ref{thm:convergence_rates} & $A\ge 4$ & {\tiny$\cO(N^2A^5(\frac{A^N}{\epsilon})^{\frac{1}{\Delta}})$} \\
    \bottomrule
  \end{tabular}
\end{table}

\subsection{Convergence time in symmetric potential games} In the following, we additionally assume that the potential game is symmetric, that is, players are interchangeable. Note that many real-world examples, such as instances of resource allocation and coverage games \cite{marden2013distributed}, are symmetric.

\begin{definition}\label{def:symmetric_game}
    A game is symmetric if for any permutation $\pi$ of $\{1,\ldots,N\}$ it holds that:
    \begin{align*}
        U_i(a_1,\ldots,a_N)=U_{\pi(i)}(a_{\pi(1)},\ldots,a_{\pi(N)}).
    \end{align*}
\end{definition}

\noindent In other words, a player's utility depends solely on the number of players selecting each action and not on their identity. Thus, in a symmetric potential game, if $A<N$, the potential function $\Phi$ can be redefined in terms of a lower-dimensional function $\Phi_{m}:\Psi^\cA\rightarrow[0,1]$, where:

\begin{align}\label{eq:shah_shin_modified_state_space}
    \Psi^\cA\eqdef \bigg\{\bigg(\frac{v_1}{N},\ldots,\frac{v_A}{N}\bigg) \: | \: v_j\in\Z_+\: \forall j\in[A],\: \sum_{j=1}^A v_j=N\bigg\}
\end{align}
with cardinality $Y=|\Psi^\cA|\le (N+1)^{A-1}$. Note that the cardinality of $\Psi^\cA$ with $\mO(N^A)$ is smaller than that of the original state space $\cAp$ with $A^N$. At the same time, for any $a\in\cAp$, it holds that $\Phi(a)=\Phi_{m}(x(a))$, where $x(a)=(x_1(a),\ldots,x_A(a))$ and $x_j(a)$ denotes
the fraction of players that selected action $j\in\cA$, \ie $x_j(a)=1/N|\{i\in[N] \:| \: a_i=j\}|$.

In the following, we assume all players follow the modified log-linear learning dynamics from \cite{shah2010dynamics}. In modified log-linear learning, a variant of log-linear learning, every player $i$ has an independent exponential clock with
rate $\alpha/z_i^t$. Player $i$'s exponential clock is an exponentially distributed random variable of mean $\alpha/z_i^t$, where $\alpha>0$ is a parameter and $z_i^t\eqdef 1/N |\{j\in[N] \: | \: a_j^t=a_i^t\}|$ counts the number of players playing the same action as player $i$. When player $i$'s clock rings, i.e., the sampled $\exp(\alpha/z_i^t)$ waiting time elapses, that player immediately resamples her action according to $p_i^t$ defined in Equation \eqref{eq:log_linear_update}. 

Modified log-linear learning induces an aperiodic and irreducible Markov chain on the lower-dimensional state space $\Psi^\cA$ with stationary distribution $\mu_m\in\Delta(\Psi^\cA)$ given by \cite[Lemma~2]{shah2010dynamics}:

\begin{align*}
    \mu_m(x) = \frac{e^{\beta\Phi_m}(x)}{\sum_{\Tx\in\Psi^\cA}e^{\beta\Phi_m}(\Tx)}, \quad\forall x\in \Psi^\cA.
\end{align*}


\begin{corollary}\label{cor:symmetric_potential}
    Consider a symmetric potential game with potential function $\Phi_m:\Psi^{\cA}\rightarrow[0,1]$. For $\epsilon\in(0,1)$ and initial distribution $\mu^0$, assume that players adhere to modified log-linear learning with:

    \begin{align*}
        \beta \ge 
        &\frac{1}{\Delta}\log\bigg((N+1)^{A-1}\bigg(\frac{1}{\epsilon Y_*}-\frac{1}{Y_*}\bigg)\bigg),
    \end{align*}
     where $\!Y_* \!=\lvert\{x^*\in\Psi^\cA\: | \:x^*\in\argmax_{x\in\Psi^\cA}\!\!\Phi_m(x)\}\rvert$ denotes the cardinality of the set of potential maximizers. Then, 
    
    \begin{align*}
        \mathbb{E}_{x\sim \mu^t}[\Phi_m(x)]\geq \max_{x\in\Psi^\cA}\Phi_m(x)-\epsilon,
    \end{align*}
    for $
         t \ge\frac{N}{\alpha c}e^{3\beta}\left(\log((A-1)\log( N +1)) + \log \beta + 2\log \frac{4}{\epsilon}\right)$, where $c>0$ is some constant. 
\end{corollary}

\noindent We provide the proof of the above corollary in Appendix \ref{app:modified_log_linear}. The result states that after {\small$t=\cO(N(N^A/\epsilon)^{\frac{1}{\Delta}})$} rounds the expected potential function value at time $t$ is $\epsilon$-optimal. The polynomial dependence on $N$ crucially relies on considering exponential clocks with dynamic means of the form $\alpha/z_i^t$ in modified log-linear learning. In contrast, in classical log-linear learning, which considers exponential clocks with mean $1$, the convergence time depends exponentially on $N$ \cite[Example~2]{shah2010dynamics}. Furthermore, if {\small$A = \mathcal{O}( \log N/ \log \log N )$}, then the convergence time to an $\epsilon$-efficient Nash equilibrium is {\small$\cO(N(N^{\log N }/\epsilon)^{\frac{1}{\Delta}})$} which is quasi-polynomial in the input size $\log (N^A)=\cO(N^{\log N})$ and polynomial in $1/\epsilon$. Note that the assumption that the cardinality of the action space is exponentially smaller than the number of players is realistic in applications like routing games, where the number of routes is considerably smaller than the number of agents.

For symmetric potential games with a $\lambda$-Lipschitz-continuous potential function, \cite{shah2010dynamics} prove a convergence time of {\small $\cO(N (\frac{A\lambda}{\epsilon})^{\frac{A}{\epsilon}})$}. Our result does not rely on this Lipschitz continuity assumption and significantly improves the dependence on $\epsilon$ from exponential to polynomial. 

\section{Proof of Theorem \ref{thm:convergence_rates}}
\label{sec:proof_main_theorem}

In this section, we prove our main result Theorem \ref{thm:convergence_rates}. As we analyze the convergence time of log-linear learning by studying the associated Markov chain, we review the basic concepts and properties of Markov chains in Appendix \ref{app:markov_chains}. 

The proof leverages the following decomposition based on the Cauchy-Schwarz inequality: 

\begin{align*}
        \mathbb{E}_{a\sim\mu^t}[\Phi(a)]
        &\geq \underbrace{\mathbb{E}_{a\sim\mu}[\Phi(a)]}_{\text{First term}} - 2\underbrace{\|\mu^t-\mu\|_{TV}}_{\text{Second term}}\underbrace{\max_{a\in\cAp}\Phi(a)}_{\le 1}.
    \end{align*}
    To control the first term, we propose a novel lemma, Lemma \ref{lem:beta_bound}, that provides a lower bound on $\mathbb{E}_{a\sim\mu}[\Phi(a)]$ if $\beta$ is sufficiently large. The second term is related to the mixing time of log-linear learning. Thus, to control the second term, we leverage mixing-time bounds based on the log-Sobolev constant, where the log-Sobolev constant is defined in Equation \eqref{eq:def_log_Sobolev} in Appendix \ref{app:markov_chains}. We rely on another novel lemma, Lemma \ref{lem:general_bound_log_Sobolev}, to bound the log-Sobolev constant of the Markov chain induced by log-linear learning. 
    Before we provide a formal proof of Theorem \ref{thm:convergence_rates}, we state the two lemmas mentioned above. 
    \begin{lemma}\label{lem:beta_bound}
    For any $\epsilon\in(0,1)$, if all players adhere to log-linear learning with:
  \begin{align*}
                \beta&\geq \frac{1}{\Delta}\log\left(\left(\Ap-A_*^N\right)\left(\frac{1}{\epsilon A_*^N}-\frac{1}{A_*^N}\right)\right),
            \end{align*}
            then it holds that $\mathbb{E}_{a\sim\mu}[\Phi(a)]\geq \max_{a\in\cAp}\Phi(a)-\epsilon.$ 
\end{lemma}
\noindent The proof of this lemma is provided in Appendix \ref{app:proof_beta_bound}. 
\begin{lemma}\label{lem:general_bound_log_Sobolev}
    Consider a Markov chain $X_t$ over state space $\cAp$ with $A\ge 4$. Assume that there exists $p_{\min},  p_{\max} \in (0,1]$, such that the corresponding transition matrix $P$ satisfies:

\begin{align}\label{eq:general_form}
         \frac{1}{N}p_{\min} \mathds{1}_{\Ta \in \cN(a)} \leq P_{a,\Ta}\leq \min\{1,\frac{1}{N}p_{\max}\} \mathds{1}_{\Ta \in \cN(a)} 
    \end{align}
    where $\cN(a) = \{\Ta\in\cAp \: | \:\exists i\in [N]: \Ta_{-i} = a_{-i}\}$. Then the log-Sobolev constant $\rho(PP^*)$ of $PP^*$ is lower bounded by:

        \begin{align*}
            \rho(PP^*)\geq \frac{16\pi^2A^{N-2}\mu_{\min} p_{\min}^3}{25N^2},
        \end{align*}
        where $\mu$ is the stationary distribution  of the Markov chain induced by $P$ and $\mu_{\min} =\min_{a\in\cAp} \mu(a)$.
\end{lemma}
\noindent The bound on $\rho(PP^*)$ applies to any Markov chain whose transition matrix satisfies Equation \eqref{eq:general_form}. In particular, it applies to the Markov chain induced by log-linear learning since the transition matrix specified in Inequality \eqref{eq:transition_prob_log_linear} satisfies Equation \eqref{eq:general_form}. As this lemma is a key technical result enabling the proof of Theorem \ref{thm:convergence_rates}, we provide a proof in Section \ref{proof:bound_log_Sobolev}.
\\
\begin{proof}(\textit{Theorem \ref{thm:convergence_rates}})
    By Cauchy-Schwarz inequality, the following holds: 

\begin{align}\label{eq:cauchy_schwarz_bound}
        \mathbb{E}_{a\sim\mu^t}[\Phi(a)]
        &\geq \underbrace{\mathbb{E}_{a\sim\mu}[\Phi(a)]}_{\text{First term}} - 2\underbrace{\|\mu^t-\mu\|_{TV}}_{\text{Second term}}\underbrace{\max_{a\in\cAp}\Phi(a)}_{\le 1}.
    \end{align}
    \textbf{First term:} If $\beta$ is set as in Lemma \ref{lem:beta_bound} replacing $\epsilon$ by $\epsilon/2$, then it holds that:
\begin{align}\label{eq:term1_bound}
    \mathbb{E}_{a\sim\mu}[\Phi(a)]\ge \max_{a\in\cAp}\Phi(a)-\epsilon/2.
\end{align}
\noindent
\textbf{Second term:} This term is related to the mixing time of log-linear learning, namely $
    \tm{P}(\epsilon) \eqdef \min\{t\in\bN \: | \: \|\mu^t-\mu\|_{TV}\leq \epsilon\}$ \cite{levin2017markov}. We control the mixing time of log-linear learning using the following mixing time bound \cite[Section~3]{diaconis1996logarithmic}:

\begin{align}\label{eq:mixing_time_bound}
    \tm{P}(\epsilon/4)\leq\frac{1}{\rho(PP^*)}\bigg(\log\log \frac{1}{\mu_{\min}}+2\log\frac{4}{\epsilon}\bigg),
\end{align}
where $\rho(PP^*)$ is the log-Sobolev constant of the Markov chain induced by the transition matrix $PP^*$, and $P^*$ is the time-reversal of $P$. Therefore, if:
\begin{equation}\label{eq:implicit_mixing_time}
    t\ge\frac{1}{\rho(PP^*)}\left(\log\log \frac{1}{\mu_{\min}}+2\log\frac{4}{\epsilon}\right),
\end{equation}
then it holds that $\|\mu^t-\mu\|_{TV}\le \epsilon/4$. By Lemma \ref{lem:general_bound_log_Sobolev}, the log-Sobolev constant $\rho(PP^*)$ can be lower-bounded as:

\begin{align}\label{eq:log-Sobolev_log_linear}
        \rho(PP^*)\geq \frac{16\pi^2 e^{-4\beta}}{25N^2 A^5},
\end{align}
where we used that by definition of $P$ and $\mu$ in Equations \eqref{eq:transition_prob_log_linear} and \eqref{eq:stationary_dist_log_linear}, respectively, $\mu_{\min}$ and $p_{\min}$ can be lower-bounded as follows:

    \begin{align}
        &\mu_{\min}=\min_{a\in\cAp}\mu(a)\geq \frac{e^{-\beta}}{\Ap}\label{eq:mu_min_log-linear}\\
        & P_{a,\Ta}\geq \frac{e^{-\beta}}{N A},\quad\forall \Ta\in\cAp(a)\ \Rightarrow\ p_{\min}=\frac{e^{-\beta}}{A}.\nonumber
    \end{align}
    Plugging Inequality \eqref{eq:log-Sobolev_log_linear} and Inequality \eqref{eq:mu_min_log-linear} into Inequality \eqref{eq:implicit_mixing_time}, it follows that if:
\begin{equation*}
    t\ge\frac{25N^2A^5}{16\pi^2}e^{4\beta}\left(\log\log \frac{\Ap}{e^{-\beta}}+2\log\frac{4}{\epsilon}\right),
\end{equation*}
then it holds that:
\begin{align}\label{eq:term2_bound}
    \|\mu^t-\mu\|_{TV}\le \epsilon/4.
\end{align}
\\
\noindent
\textbf{Combination:} If $\beta$ is set as in Lemma \ref{lem:beta_bound} replacing $\epsilon$ by $\epsilon/2$ and $t\ge\frac{25N^2A^5}{16\pi^2}e^{4\beta}\left(\log\log \frac{\Ap}{e^{-\beta}}+2\log\frac{4}{\epsilon}\right)$, then it holds that:

\begin{align*}
        \mathbb{E}_{a\sim\mu^t}[\Phi(a)]
        &\overset{(i)}{\geq} \mathbb{E}_{a\sim\mu}[\Phi(a)] - 2\|\mu^t-\mu\|_{TV} \max_{a\in\cAp}\Phi(a)\\
        &\overset{(ii)}{\geq} \max_{a\in\cAp} \Phi(a)-\frac{\epsilon}{2}-\frac{2\epsilon}{4},\\
        &= \max_{a\in\cAp} \Phi(a)-\epsilon,
    \end{align*}
    where in (i) we used Inequality \eqref{eq:cauchy_schwarz_bound} and (ii) follows from Inequality \eqref{eq:term1_bound}, Inequality \eqref{eq:term2_bound}, and the fact that $\Phi(\cdot)\in[0,1]$. This concludes the proof. 
\end{proof}

\subsection{Proof of Lemma \ref{lem:general_bound_log_Sobolev}:}\label{proof:bound_log_Sobolev}
The general idea of the proof is to lower bound the log-Sobolev constant of the Markov chain $X_t$ with transition matrix $P$ given by Inequality \eqref{eq:general_form}, in terms of the log-Sobolev constant of another Markov chain, for which a lower bound is known. In particular, we make use of the following lemma.

\begin{lemma}{\cite[Corollary~2.15]{montenegro2006mathematical}}\label{lem:cor2.15}
Consider two Markov chains $X_t$ and $\hat{X}_t$ defined on the same state space with transition matrix $P$ and $\hat{P}$, respectively, and stationary distribution $\mu$ and $\hat{\mu}$, respectively. Then, the log-Sobolev constant $\rho(P)$ of Markov chain $X_t$ is lower-bounded as follows:

    \begin{align*}
        \rho(P)\geq\frac{1}{MC}\rho(\hat{P}),
    \end{align*}
    where $
    M=\max_{a\in\cAp}\frac{\mu(a)}{\hat{\mu}(a)}$ and $C =\max_{a\neq\Tilde{a}: (P)_{a,\Tilde{a}}\neq 0} \frac{\hat{\mu}(a)\hat{P}_{a,\Tilde{a}}}{\mu(a)(P)_{a,\Tilde{a}}}$.
    
\end{lemma}
\ \\
\\
\begin{proof}(\textit{Lemma \ref{lem:general_bound_log_Sobolev})}


\noindent Consider a Markov chain $X_t^*$ with transition matrix $PP^*$. Let $X_t$ be the Markov chain with transition matrix $P$ defined in Inequality \eqref{eq:general_form}. We will use Lemma \ref{lem:cor2.15} to obtain a lower bound on the log-Sobolev constant $\rho(PP^*)$ in terms of $\rho(P)$.

\paragraph{Comparison of $X_t^*$ and $X_t$:} Note that $X_t$ is aperiodic and irreducible and thus a unique stationary distribution $\mu$ exists with $\mu^t=\mu^0P^t\rightarrow\mu$ for $t\rightarrow\infty$, where $\mu^0$ is any initial distribution. Furthermore, $\mu_{\min}>0$ follows from the irreducibility of $X_t$. 

The Markov chain $X^*_t$ is also aperiodic and irreducible since $X_t$ is aperiodic and irreducible. Concretely, since $P$ contains self-loops, \ie $P_{a,a}>0$, it follows that $PP^*$ contains self-loops:

\begin{align*}
    (PP^*)_{a,a}&= \sum_{a'\in\mA}P_{a,a'}P^*_{a',a}= \sum_{a'\in\mA}P_{a,a'}\frac{\mu(a)P_{a,a'}}{\mu(a')}\\
    &\geq P_{a,a}P_{a,a}>0,
\end{align*}
and thus $X^*_t$ is aperiodic. Furthermore, for any $a,\Ta\in\cAp$:

\begin{align*}
    (PP^*)^N_{a,\Ta}&=\sum_{\substack{a_l\in\cAp\\l=1,\ldots,N-1}}(PP^*)_{a,a_1}\ldots(PP^*)_{a_{N-1},\Ta}\\
    &=\sum_{\substack{a_l\in\cAp\\l=1,\ldots,N-1}}\sum_{a'\in\cAp}P_{a,a'}P^*_{a',a_1}\ldots\sum_{a'\in\cAp}P_{a_{N-1},a'}P^*_{a',\Ta}\\
    &\geq\sum_{\substack{a_l\in\cAp\\l=1,\ldots,N-1}}P_{a,a_1}P_{a_1,a_1}\ldots P_{a_{N-1},\Ta}P_{\Ta,\Ta}>0,
\end{align*}
where we used that $P^N_{a,\Ta}>0$ and $P_{a,a}>0$ for all $a,\Ta\in\cAp$ as well as the identity $\mu(a)P^*_{a,\Ta}=\mu(\Ta)P_{\Ta,a}$. It follows that $X^*_t$ is irreducible. Thus, for $X^*_t$, a unique stationary distribution exists. More specifically $\mu$ is the stationary distribution of $PP^*$ since by \cite[Proposition~1.23]{levin2017markov} the stationary distribution of $P^*$ is given by $\mu$ and since $\mu PP^*=\mu P^*=\mu$. Furthermore, the following holds for the transition matrix $PP^*$:

\begin{align*}
  \frac{1}{N}p_{\min}^2\mathds{1}_{\Ta \in \cAp(a)}
\leq (PP^*)_{a,\Ta}
    \le\mathds{1}_{\Ta \in \cAp(a)}
\end{align*}
where 

\begin{align*}
    (PP^*)_{a,\Ta}=\sum_{a'\in\mA}P_{a,a'}P^*_{a',\Ta}\geq P_{a,\Ta}P^*_{\Ta,\Ta}\geq P_{a,\Ta}P^*_{\Ta,\Ta}\geq \frac{p_{\min}^2}{N}.
\end{align*}
Now, we apply Lemma \ref{lem:cor2.15} to lower-bound the log-Soblev constant $\rho(PP^*)$ of $X_t^*$ in terms of the log-Sobolev constant $\rho(P)$ of $X_t$ as follows:

    \begin{align}\label{eq:term1_general}
        \rho(PP^*)\geq\frac{1}{MC}\rho(P)\geq \frac{ p_{\min}^2}{N} \rho(P).
    \end{align}
    where

    \begin{align*}
    &M=\max_{a\in\cAp}\frac{\mu(a)}{\mu(a)}=1\\
    &C =\max_{a\neq\Tilde{a}: (PP^*)_{a,\Tilde{a}}\neq 0} \frac{\mu(a)P_{a,\Tilde{a}}}{\mu(a)(PP^*)_{a,\Tilde{a}}}\leq  \frac{N}{ p_{\min}^2}.
    \end{align*}
    Next, we consider the Markov chain $\hX_t$ with transition matrix $\hP$ specified as $\hP_{a,\Ta}=\frac{1}{NA}\mathds{1}_{\Ta \in \cN(a)}$, where $\cN(a) = \{\Ta\in\cAp \: | \:\exists i\in [N]: \Ta_{-i} = a_{-i}\}$.

\paragraph{Comparison of $X_t$ and $\hX_t$:} Note that $\hX_t$ is aperiodic and irreducible with stationary distribution $\hmu(a)=1/\Ap$. This can be verified by checking the detailed balance equations given by $\hmu(a)\hP_{a,\Ta}=\hmu(\Ta)\hP_{\Ta,a}$ for all $a,\Ta\in\Ap$. 

Next, we use \cite[Corollary~2.15]{montenegro2006mathematical} to lower-bound the log-Soblev constant $\rho(P)$ of $X_t$ in terms of the log-Sobolev constant $\rho(\hP)$ of $\hX_t$. To this end, we compute $M$ and $C$ of the Markov chains $X_t$ and $\hX_t$:

    \begin{align*}
        &M=\max_{a\in\cAp}\frac{\mu(a)}{\hmu(a)}\leq \Ap\\
    &C = \max_{a\neq\Tilde{a}: P_{a,\Tilde{a}}\neq 0} \frac{\hmu(a)\hP_{a,\Tilde{a}}}{\mu(a)P_{a,\Tilde{a}}}
    \leq  \frac{N}{\Ap NA\mu_{\min}p_{\min}},
    \end{align*}
    Thus, the log-Soblev constant $\rho(P)$ can be lower-bounded by:

\begin{align}\label{eq:term2_general}
        \rho(P)\geq \frac{1}{MC}\rho(\hat{P})\geq \Ap A\mu_{\min}p_{\min}\rho(\hat{P}).
    \end{align}
    Lastly, we consider the product chain $\bX_t$ with $\bX_t= \prod_{i=1}^N\bX_{i_t}$ on the state space $\mathbb{Z}_{\Kp}=\prod_{i=1}^N\mathbb{Z}_{K}$ with $\Z_{K}=\{1,\ldots,K\}$ and  $K\geq 4$.

\paragraph{Comparison of $\hX_t$ and $\bX_t$:}  Here, each $\{\bX_{i,t}\}_{t\in\N}$ is a simple random walk on $\Z_{K}$ with transition matrix $\bP_{i_{k,k\pm 1}}$ specified as:
    \begin{align*}
        &\bP_{i,(k,k\pm 1)}= 1/2 \qquad \text{ for } 2\leq k\leq K-1,\\
        &\bP_{i,(K,1)} = \bP_{i,(K,K-1)} = 1/2, \\
        &\bP_{i,(1,2)} = \bP_{i,(1,K)} = 1/2,
    \end{align*}

\noindent and the stationary distribution $\bmu_i(k)$ of the simple random walk $\bX_{i,t}$ is given by: 
    \begin{align*}
        \bmu_i(k) = \frac{1}{K}, \quad\forall i\in\mathcal{N}.
    \end{align*}

  \noindent Thus, the product chain $\bX_t$ has the following transition matrix \cite[Sec.~2.5]{diaconis1996logarithmic}:

    \begin{align*}
        \bP_{\textbf{k},\Tilde{\textbf{k}}}=
\frac{1}{2N}\mathds{1}_{\Tilde{\textbf{k}}=(k_i\pm 1,\textbf{k}_{-i})},
    \end{align*}
    and the stationary distribution:

\begin{align*}
    \bmu(\textbf{k})=\prod_{i=1}^N \bmu_i(k_i) =\prod_{i=1}^N \frac{1}{K}=\frac{1}{\Kp}.
\end{align*}
Note that there is a one-to-one mapping between the set $\cA$ and the set $\mathbb{Z}_{K}$ with $|\cA|=A=K$ and thus a one-to-one mapping between the set $\cAp$ and the set $\mathbb{Z}_{\Kp}$ with  with $\Ap=K^N$. Therefore, we can assume that the Markov chains $\hX_t$ and $\bX_t$ operate on the same state space. To this end, we compute $M$ and $C$ of the Markov chains $\hX_t$ and $\bX_t$:

\begin{align*}
    &M=\max_{a\in\cAp}\frac{\hmu(a)}{\bmu(a)}=\frac{\Ap}{\Ap}=1\\
    &C = \max_{a\neq\Tilde{a}: \hP_{a,\Tilde{a}}\neq 0} \frac{\bmu(a)\bP_{a,\Tilde{a}}}{\hmu(a)\hP_{a,\Tilde{a}}}=\frac{A}{2}.
    \end{align*}
    Thus, the log-Soblev constant $\rho(\hP)$ can be lower-bounded:

\begin{align*}
    \rho(\hP)\geq\frac{1}{MC}\rho(\bP)\geq \frac{2}{A} \rho(\bP).
\end{align*}
For the simple random walk $\bX_{i,t}$ a bound on the log-Sobolev constant $\rho(\bP_i)$ is known with $\rho(\bP_i)\geq \frac{8\pi^2}{25 K^2}$ \cite[Example~4.2]{diaconis1996logarithmic}. Then, by  \cite[Lemma~3.2]{diaconis1996logarithmic}, the log-Soblev constant $\rho(\bP)$ of the product chain $\bX_t$ is lower bounded by:

\begin{align*}
    \rho(\bP)=\frac{1}{N}\min_{i\in\{1,\ldots,N\}}\rho(\bP_i)\geq   \frac{8\pi^2}{25NK^2}.
\end{align*}
Thus, $\rho(\hP)$ can be lower-bounded by:

\begin{align}\label{eq:term3_general}
        \rho(\hP)\geq \frac{2}{A}\rho(\bP)\geq \frac{16\pi^2}{25NA^3}.
    \end{align}
    
\paragraph{Combination:} Combining Equations \eqref{eq:term1_general}, \eqref{eq:term2_general}, and \eqref{eq:term3_general}, we conclude that the log-Sobolev constant $\rho(PP^*)$ is lower-bounded by:

    \begin{align*}
        \rho(PP^*)\geq \frac{16\pi^2\Ap\mu_{\min} p_{\min}^3}{25N^2A^2}.
    \end{align*}
    This concludes the proof of Lemma \ref{lem:general_bound_log_Sobolev}.

    
\end{proof}

\section{Robustness of log-linear learning}

In this section, we show a convergence time guarantee in settings where players have less feedback information, Subsection \ref{sec:binary_loglinear}, and in settings where log-linear learning is subject to perturbations such as noisy utility observations, Subsection \ref{sec:perturbed_learning}.

\subsection{Reduced feedback}
\label{sec:binary_loglinear}

Log-linear learning requires players to observe their utilities for all possible actions given the other players' actions. Having such full-information feedback when
action sets are large can be demanding. Binary log-linear learning \cite{arslan2007autonomous, marden2007connections} alleviates this limitation by requiring two-point feedback, reducing the needed feedback by a factor $A$ per round. 
We briefly review the binary log-linear learning rule. 


Binary log-linear learning proceeds as log-linear learning with the distinction that the player $i$ allowed to alter her action first samples a trial action $\Ta_i$ uniformly from her action set $\cA$. She then plays according to the strategy:

\begin{align*}
     &p_i^t(a_i)=\begin{cases}
         \frac{e^{\beta U_i(a_i,a_{-i}^{t-1})}}{e^{\beta U_i(a_i^{t-1},a_{-i}^{t-1})} + e^{\beta U_i(\Ta_i,a_{-i}^{t-1})}}, & \text{for $a_i\in\{\Ta_i,a_i^{t-1}\}$.}\\
         0, & \text{otherwise.}
     \end{cases}
\end{align*} 
Here, player $i$ can either repeat her action $a_i^{t-1}$ or play one other randomly sampled action $\Ta_i$ rather than any action $a_i\in\cA$ as in log-linear learning. Next, we derive the first finite-time convergence bound of binary log-linear learning to an $\epsilon$-efficient Nash equilibrium.

\begin{theorem}\label{thm:convergence_rates_binary}
    Consider a potential game with potential function $\Phi:\cAp\rightarrow [0,1]$ and $A\ge 4$. For $\epsilon\in(0,1)$ and initial distribution $\mu^0$, assume that players adhere to binary log-linear learning with $\beta = \Omega\left(\frac{1}{\Delta}\log \frac{\cAp}{\epsilon}\right)$.
    Then, it holds that $\mathbb{E}_{a\sim \mu^t}[\Phi(a)]\geq \max_{a\in\cAp}\Phi(a)-\epsilon$ for

    \begin{align*}
        t&\geq \frac{25N^2 A^5}{2\pi^2 }e^{4\beta}\left(\log\log \Ap+\log \beta+2\log\frac{4}{\epsilon}\right)\\
        &\approx \cO\left(N^2 A^5\left(A^N/\epsilon\right)^{\frac{N}{\Delta}}\right).
    \end{align*}    
\end{theorem}

\noindent The proof follows similar arguments as that of Theorem \ref{thm:convergence_rates} and provide a detailed proof in Appendix \ref{app:thm2}.
 We remark that with significantly less feedback per round, binary log-linear achieves the same convergence speed as log-linear learning up to a factor of $8$. 

\subsection{Perturbed log-linear learning}\label{sec:perturbed_learning}

Classical log-linear learning relies on two limiting assumptions: 1) Players have access to their exact utilities. However, in real-world applications, the presence of noise is typical as uncertainties and hidden factors generate inexact measurements. 2) Players are rational. However, empirical evidence suggests that players have limited rationality and therefore may occasionally deviate from the log-linear learning rule in practical scenarios. Our next result generalizes Theorem \ref{thm:convergence_rates} to the case where the log-linear learning rule is subject to small perturbations. This generalization can address utilities with corrupted noise and log-learning learning mixed with uniform exploration, as will be shown.


\hide{
We first derive a tight Lipschitz constant for the known result regarding the Lipschitz-continuity of stationary distributions of Markov chains as a function of their transition matrix \cite{zhang2023global}. We then leverage this result to prove our main result on the convergence time of perturbed log-linear learning to an $\epsilon$-efficient NE.

\begin{lemma}[Lipschitzness] \label{lem:Lipschitz}
    Consider two irreducible and aperiodic transition matrices $P_1, P_2 \in\R^{\Ap\times\Ap}$. Let $\mu_1$ and $\mu_2$ be the stationary distributions of the Markov chains induced by $P_1$ and $P_2$, respectively. Then, the following holds:
    
    \begin{equation*}
        \|\mu_1-\mu_2\|_2 \le \min\{L(P_1),L(P_2)\} \|P_1-P_2\|_2,
    \end{equation*}
   where {\small$L(P_k) \eqdef \frac{2 \Ap}{\rho(P_k P_k^*)} (\log\log \frac{1}{\mu_{k,\min}}+\log(8 \Ap))$} and $\mu_{k,\min} = \min_{a\in\cAp}\mu_k(a)$ for $k=1,2$.    
\end{lemma}
We provide a proof in Appendix \ref{app:lemma_lipschitz}. We considerably improve the Lipschitz constant by using the mixing time bound based on log-Sobolev inequalities (Lemmas \ref{lem:general_mixing_time} and \ref{lem:general_bound_log_Sobolev}). In particular, compared to \cite[Lemma 24]{zhang2023global} which entails a Lipschitz constant $L=\Tilde{\mathcal{O}}((e/p_{\min})^N)$ our Lemma \ref{lem:Lipschitz} implies that $L = \Tilde{\mathcal{O}}(1/(\mu_{\min} p_{\min}^3))$.\footnote{This can be seen by injecting the log-Sobolev bound of lemma \ref{lem:general_bound_log_Sobolev} into our Lemma above.} Now, we state the main result of this section.
}

\begin{theorem}\label{thm:General_Convergence}
    Consider a potential game with a potential function $\Phi:\cAp\rightarrow[0,1]$ and $A\ge4$. Let $P_{\ell}$ denote the transition matrix of log-linear learning and $L$ denote a Lipschitz constant of order {\small$\mathcal{\Tilde{O}}\left(N^2 A^{N+5} e^{\log(\Ap /\epsilon)/\Delta} \right)$}. Furthermore, consider a learning rule with transition matrix $P$ such that there exists $p_{\min},  p_{\max} \in (0,1]$, with:
    
\begin{align}\label{eq:pertrubed_transitions}
        \frac{1}{N}p_{\min} \mathds{1}_{\Ta \in \cN(a)} \leq P_{a,\Ta}\leq \min\{1,\frac{1}{N}p_{\max}\} \mathds{1}_{\Ta \in \cN(a)}
    \end{align}
    for all $a, \Ta \in \cA$. For $\epsilon\in(0,1)$ and initial distribution $\mu^0$, assume all players adhere to this learning rule with {\small$\beta = \Tilde{\Omega}\left(\frac{1}{\Delta}\log \frac{A^N}{\epsilon}\right)$}. 
    Then,
    
    \begin{equation*}
        \bE_{a\sim \mu^t}[\Phi(a)] \ge \max_{a\in \cA} \Phi(a) - \epsilon - L\sqrt{\Ap} \|P-P_{\ell}\|_2,
    \end{equation*}
    for
    
    \begin{equation*}
        t \ge \frac{25 N^{3/2} e^N}{(2\pi)^{5/2} \Ap p_{\min}^{N+3}} \log\left(\frac{4 \Ap}{\epsilon^2} \log\frac{e^N}{p_{\min}^N \sqrt{2\pi N}}\right).
    \end{equation*}

    
\end{theorem}
\ \\
Theorem \ref{thm:General_Convergence} proves finite time convergence guarantees to reach an $\epsilon$-efficient Nash equilibrium when the stationary distribution of the learning rule is unknown assuming only that the learning rules' transition matrix $P$ is sufficiently close to the transition matrix $P_l$ induced by log-linear learning, i.e., $\|P-P_{\ell}\|_2 = \mathcal{O}(\epsilon/(L\sqrt{\Ap}))$. However, due to the unavailability of the stationary distribution of the perturbed learning rule, the convergence time is $(N/p_{\min})^N / N!$ times slower compared to log-linear learning. 

Before proving Theorem \ref{thm:General_Convergence}, we state a lemma that we will use in the proof. In the lemma, we derive a tight Lipschitz constant for the stationary distributions of Markov chains as a function of their transition matrices.

\begin{lemma}[Lipschitzness] \label{lem:Lipschitz}
        Consider two irreducible and aperiodic transition matrices $P_1, P_2 \in\R^{\Ap\times\Ap}$ with $\mu_1$ and $\mu_2$ as the stationary distributions of the Markov chains induced by $P_1$ and $P_2$, respectively. Then, the following holds:
        
        \begin{equation*}
            \|\mu_1-\mu_2\|_2 \le \min\{L(P_1),L(P_2)\} \|P_1-P_2\|_2,
        \end{equation*}
        where {\small$L(P_k) \eqdef \frac{2 \Ap}{\rho(P_k P_k^*)} (\log\log \frac{1}{\mu_{k,\min}}+\log(8 \Ap))$} and $\mu_{k,\min} = \min_{a\in\cAp}\mu_k(a)$ for $k=1,2$.    
    \end{lemma}
    
    \noindent We provide a proof of this lemma in Appendix \ref{app:robustness}. 
    Compared to \cite[Lemma 24]{zhang2023global} which entails a Lipschitz constant $L=\Tilde{\mathcal{O}}((e/p_{\min})^N)$, Lemma \ref{lem:Lipschitz} improves the Lipschitz constant to $L = \Tilde{\mathcal{O}}(1/(\mu_{\min} p_{\min}^3))$ leveraging mixing-time bounds based on the log-Sobolev constant.
%
%
%
\hide{
We first derive a tight Lipschitz constant for the known result regarding the Lipschitz-continuity of stationary distributions of Markov chains as a function of their transition matrix \cite{zhang2023global}. We then leverage this result to prove our main result on the convergence time of perturbed log-linear learning to an $\epsilon$-efficient NE.

\begin{lemma}[Lipschitzness] \label{lem:Lipschitz}
    Consider two irreducible and aperiodic transition matrices $P_1, P_2 \in\R^{\Ap\times\Ap}$. Let $\mu_1$ and $\mu_2$ be the stationary distributions of the Markov chains induced by $P_1$ and $P_2$, respectively. Then, the following holds:
    
    \begin{equation*}
        \|\mu_1-\mu_2\|_2 \le \min\{L(P_1),L(P_2)\} \|P_1-P_2\|_2,
    \end{equation*}
    where {\small$L(P_k) \eqdef \frac{2 \Ap}{\rho(P_k P_k^*)} (\log\log \frac{1}{\mu_{k,\min}}+\log(8 \Ap))$} and $\mu_{k,\min} = \min_{a\in\cAp}\mu_k(a)$ for $k=1,2$.    
\end{lemma}
We provide a proof in Section \ref{sec:lemma_lipschitz}. We considerably improve the Lipschitz constant by using the mixing time bound based on log-Sobolev inequalities (Lemmas \ref{lem:general_mixing_time} and \ref{lem:general_bound_log_Sobolev}). In particular, compared to \cite[Lemma 24]{zhang2023global} which entails a Lipschitz constant $L=\Tilde{\mathcal{O}}((e/p_{\min})^N)$ our Lemma \ref{lem:Lipschitz} implies that $L = \Tilde{\mathcal{O}}(1/(\mu_{\min} p_{\min}^3))$.\footnote{This can be seen by injecting the log-Sobolev bound of lemma \ref{lem:general_bound_log_Sobolev} into our Lemma above.} Now, we state the main result of this section.}
Next, we provide a proof of Theorem \ref{thm:General_Convergence}.
\\
\begin{proof}(\textit{Theorem \ref{thm:General_Convergence}})
    Consider a learning rule with transition matrix $P$ satisfying Equation~\eqref{eq:pertrubed_transitions}. We first provide a decomposition that relates the expected value of the potential when the agents follow $P$ to the same quantity where the agents instead follow $P_\ell$ defined in Equation \eqref{eq:transition_prob_log_linear}.  We have for all $t, t' \in \bN$ that:
    
    \begin{align}
        &\bE_{a\sim\mu_0 P^t}[\Phi(a)] \nonumber\\
        &= \bE_{a\sim\mu_0 P_{\ell}^{t'}}[\Phi(a)] + \bE_{a\sim\mu_0 P^t}[\Phi(a)] - \bE_{a\sim\mu_0 P_{\ell}^{t'}}[\Phi(a)] \nonumber\\
        &\ge \bE_{a\sim\mu_0 P_{\ell}^{t'}}[\Phi(a)] - \sqrt{\Ap} \|P^t - P_{\ell}^{t'}\|_2  \label{ineq:Decomposition}
    \end{align}
    where we used that $|\Phi(a)|\le 1$ for all $a\in \cAp$ and $\|\cdot\|_1 \le \sqrt{\Ap}\|\cdot\|_2$. 

\paragraph{Decomposition:} We start with the following decomposition:
    \begin{align}\label{eq:Decomposition_Robustness}
        \|P^t - P_{\ell}^{t'}\|_2 &\le \|P^t - \mu \|_2 + \|P_{\ell}^{t'} - \mu_{\ell}\|_2 + \|\mu - \mu_{\ell}\|_2\\
        &\le \|P^t - \mu \|_2 + \|P_{\ell}^{t'} - \mu_{\ell}\|_2 + L(P_{\ell}) \|P-P_{\ell}\|_2\nonumber\\
        &\le 2\|P^t - \mu \|_{TV} + \|P_{\ell}^{t'} - \mu_{\ell}\|_2 + L(P_{\ell}) \|P-P_{\ell}\|_2\nonumber
    \end{align}
    where we used Lemma \ref{lem:Lipschitz} in the second inequality. In Theorem \ref{thm:convergence_rates}, we showed that $\mu_{\ell, \min} \geq \frac{e^{-\beta}}{\Ap}$ and $\rho(P_\ell P_\ell^*)\geq \frac{16\pi^2 e^{-4\beta}}{25N^2 A^5}$, therefore 
    {\small
    $
        L(P_\ell) \le \frac{25 N^2 A^{N+5} e^{4 \beta }}{8 \pi^2} (\log\log \Ap e^\beta +\log(8 \Ap))$
    }.
    
    \noindent Under the following three conditions:
    
    \begin{itemize}
        \item $t \ge  \tm{P}(\epsilon/(4\sqrt{\Ap})))$,
        \item $t' \to \infty$,
        \item $\beta = \frac{1}{\Delta} \log\left(\left(\Ap-A_*^N\right)\left(\frac{4}{\epsilon A_*^N}-\frac{1}{A_*^N}\right)\right)$,
    \end{itemize}
    we establish the following three inequalities:
    
    \begin{enumerate}
        \item $\|P^t - P_{\ell}^t\|_2 \le \epsilon / \left(2\sqrt{\Ap}\right) + L(P_{\ell}) \|P-P_{\ell}\|_2$,
        \item $\bE_{a\sim\mu^0P_{\ell}^{t'}}[\Phi(a)] \ge \max_{a\in \cAp} \Phi(a) - \epsilon/2$,
        \item $L(P_\ell) = \mathcal{O}\Big( N^2 A^{N+5} e^{\frac{\log(\Ap/\epsilon)}{\Delta}} \Big(\log\log \Ap e^{\frac{\log(\Ap/\epsilon)}\Delta}  +\log(\Ap) \Big)\Big)$,
    \end{enumerate}
    where the second line follows from Theorem \ref{thm:convergence_rates}. Plugging the above inequalities into the decomposition \eqref{ineq:Decomposition} proves the desired result for $t \ge \tm{P}(\epsilon/(4\sqrt{\Ap}))$. 
    \\
    \noindent
    We now provide a bound on the mixing time $\tm{P}(\epsilon/(4\sqrt{\Ap}))$ governing the first term in Equation \eqref{eq:Decomposition_Robustness}. To bound the mixing time we use Inequality \eqref{eq:mixing_time_bound} and Lemma \ref{lem:general_bound_log_Sobolev}. Assuming a lower bound of $p_{\min}/N$ on the probabilities of all feasible transitions implies a lower bound on the stationary distribution as we show next.  
\\
\paragraph{Lower bound $(\mu_P)_{\min}$:} Since $P$ has a positive probability of transitioning from $a\in \cAp$ to any $\Ta \in \cN(a)$, it follows that the corresponding $N$-step transition $P^N$ has a positive probability of transitioning from any $a\in \cAp$ to any $a' \in \cAp$, \ie
    
    \begin{equation*}
        \forall a,a' \in \cAp: \quad P^N_{a,a'} \ge N! \:(p_{\min}/N)^N.
    \end{equation*}
    The least probable transitions are such that $\forall i \in [N]: \: a_i \neq a'_i$. For such transitions, the possible paths using $P^N$ are the $N!$ permutations of $\{1,\ldots,N\}$ (each of the $N$ steps is a new player updating their action) and each player $i \in [N]$ can update $a_i$ to $a'_i$ with probability larger than $p_{\min}/N$. 

    Since $P$ is an irreducible and aperiodic transition matrix, the Markov chain induced by $P$ has a unique stationary distribution $\mu_P$. It is known that the Markov chain induced by $P^N$ has the same stationary distribution $\mu$. Therefore, we have for all $a\in\cAp$:
    
    \begin{align*}
        \mu_P (a) &= \sum_{\Ta \in \cAp} P^N_{\Ta,a} \mu_P (\Ta)\\
        &\ge \sum_{\Ta \in \cAp} N! \:(p_{\min}/N)^N \mu_P (\Ta) = N! \:(p_{\min}/N)^N
    \end{align*}
    and $(\mu_P)_{\min}\ge N! \:(p_{\min}/N)^N$.

\paragraph{Deducing the mixing-time bound:} We now give an explicit bound on the mixing time of $P$. First, by Lemma \ref{lem:general_bound_log_Sobolev} have :
 
    \begin{align*}
        \rho(PP^*) &\ge \frac{16 \pi^2 \Ap (\mu_P)_{\min} p_{\min}^3}{25 N^2}
        &\ge \frac{4\pi^2 \Ap p_{\min}^{N+3} N!}{25N^{N+2}}.
    \end{align*}
    Using Stirling's formula, we have $N! \ge \sqrt{2 \pi N}\left(\frac{N}{e}\right)^N $, thus:

    \begin{align*}
        \rho(PP^*) \ge \frac{(2\pi)^{5/2} \Ap p_{\min}^{N+3}}{25 N^{3/2} e^N}.
    \end{align*}
    To conclude the proof, using Inequality \eqref{eq:mixing_time_bound} we obtain:
    
    \begin{align*}
        &\tm{P}(\epsilon/(4\sqrt{\Ap}))\\
        &\le \frac{1}{\rho(PP^*)}\left(\log\log \frac{1}{(\mu_P)_{\min}} + 2 \log\frac{4\sqrt{\Ap}}{\epsilon}\right)\\
        &\le \frac{25 N^{3/2} e^N}{(2\pi)^{5/2} \Ap p_{\min}^{N+3}} \left(\log\log\frac{e^N}{p_{\min}^N \sqrt{2\pi N}} + 2 \log\frac{4 \sqrt{\Ap}}{\epsilon}\right).
    \end{align*}
\end{proof}

We now consider two explicit types of perturbations: noisy utilities and a modified learning rule. 

\subsubsection{Corrupted utilities with additive noise} 

We assume that players observe noise-corrupted utilities $(\hat{U}_i)_{i\in[N]}$ satisfying:

\begin{equation}\label{eq:corrupted_utility}
    \hat{U}_i(a_i,a_{-i}) = U_i(a_i,a_{-i}) + \xi_i(a_i,a_{-i}), \quad \forall (a_i,a_{-i}) \in \cAp
\end{equation}
where $\xi_i(a_i,a_{-i}) \in [-\xi,\xi]$ is a bounded noise. Alternatively, the noise could be assumed to be centered i.i.d. random variables with bounded variance \cite{leslie2011equilibrium}. Using Theorem \ref{thm:General_Convergence}, we show that log-linear learning is robust to noisy feedback.

\begin{corollary}\label{cor:noise_corruption}
    Consider the setting of Theorem \ref{thm:General_Convergence} with noise-corrupted utilities as in Equation \eqref{eq:corrupted_utility}. If all players adhere to log-linear learning with $\beta = \Tilde{\Omega}\left(\frac{1}{\Delta}\log \frac{A^N}{\epsilon}\right)$ and $\xi \le 1/(2\beta)$, then
    
    \begin{equation*}
        \bE_{a\sim \mu^t}[\Phi(a)] \ge \max_{a\in \cA} \Phi(a) - \epsilon - \frac{7 L A^{3N/2}}{2 N} \beta \xi,
    \end{equation*}
    for {\small$t=\mO\left( N^{3/2} A^3 e^{N + \beta(1+2\xi)(N+3)} \log\frac{1}{\epsilon^2} \right)$} with {\small$L = \mathcal{\Tilde{O}}\left(N^2 A^{N+5} e^{\log(\Ap /\epsilon)\Delta} \right)$}. 
\end{corollary}


\noindent The proof follows from applying Theorem \ref{thm:General_Convergence} and is provided in Appendix \ref{app:cor_corruption}. Corollary \ref{cor:noise_corruption} shows that log-linear learning with corrupted utilities converges to an $\epsilon$-efficient Nash equilibrium in time polynomial in $1/\epsilon$ if the corruption magnitude $\xi$ is sufficiently small. Our finite-time convergence result extends previous works on robust learning which provide asymptotic guarantees \cite{leslie2011equilibrium,lim2013robustness,bravo2017robustness}.
The key to this result lies in showing that the transition matrix of the Markov chain induced by corrupted utilities is close to its corruption-free counterpart. 

\begin{figure*}[!h]
\centering
\begin{subfigure}[b]{0.31\textwidth}
   \includegraphics[width=1\linewidth]{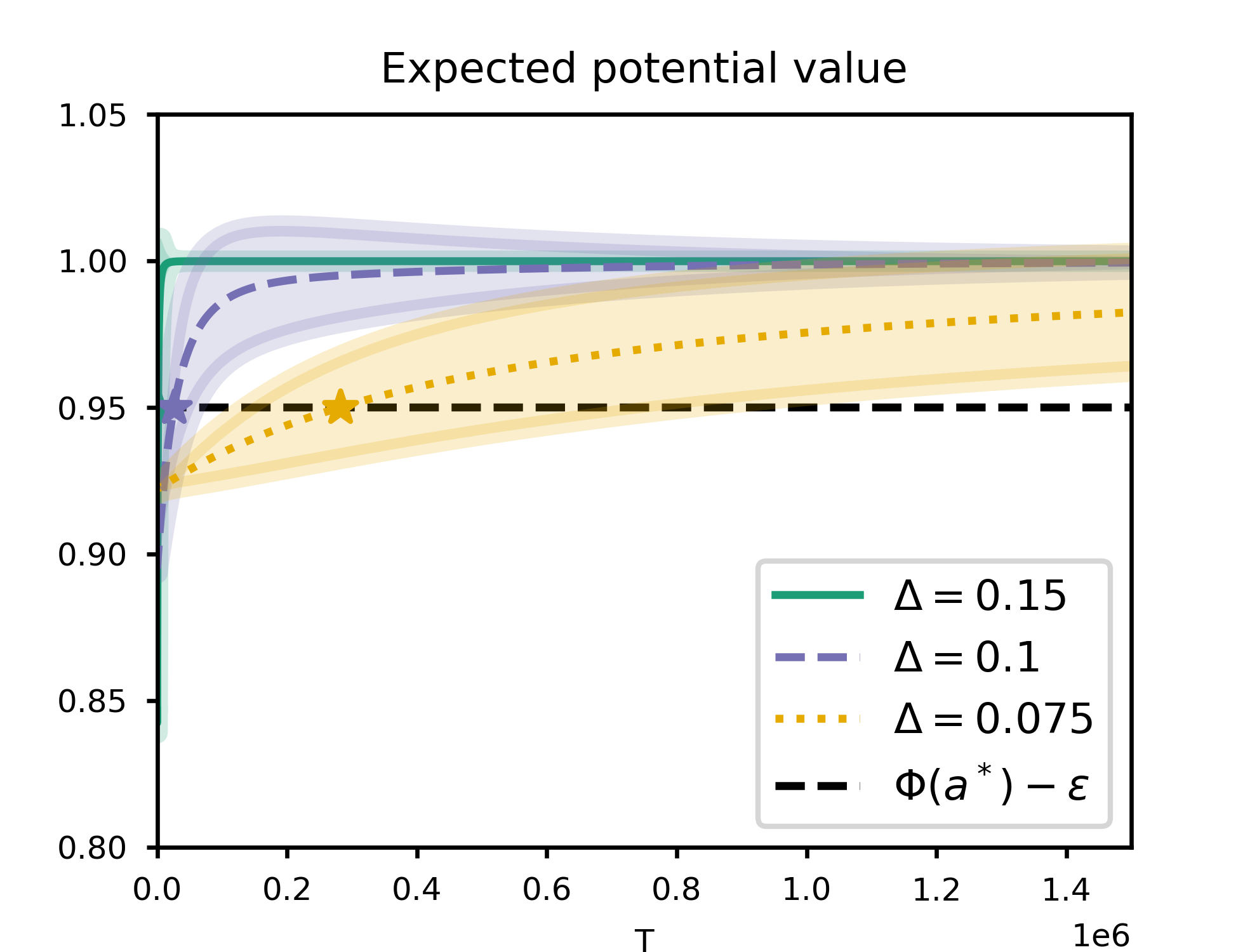}
\end{subfigure}
\begin{subfigure}[b]{0.31\textwidth}
   \includegraphics[width=1\linewidth]{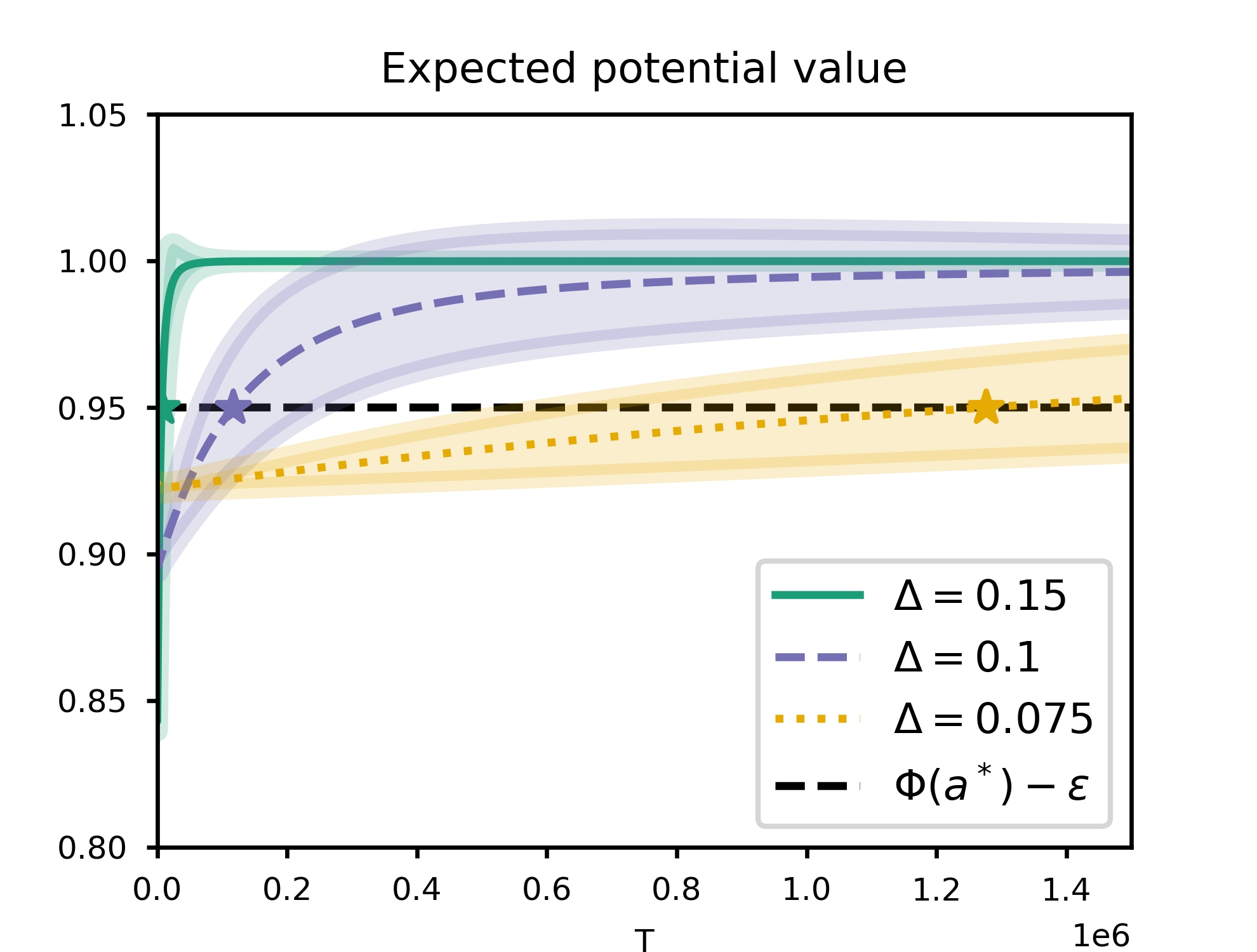}
\end{subfigure}
\begin{subfigure}[b]{0.31\textwidth}
   \includegraphics[width=1\linewidth]{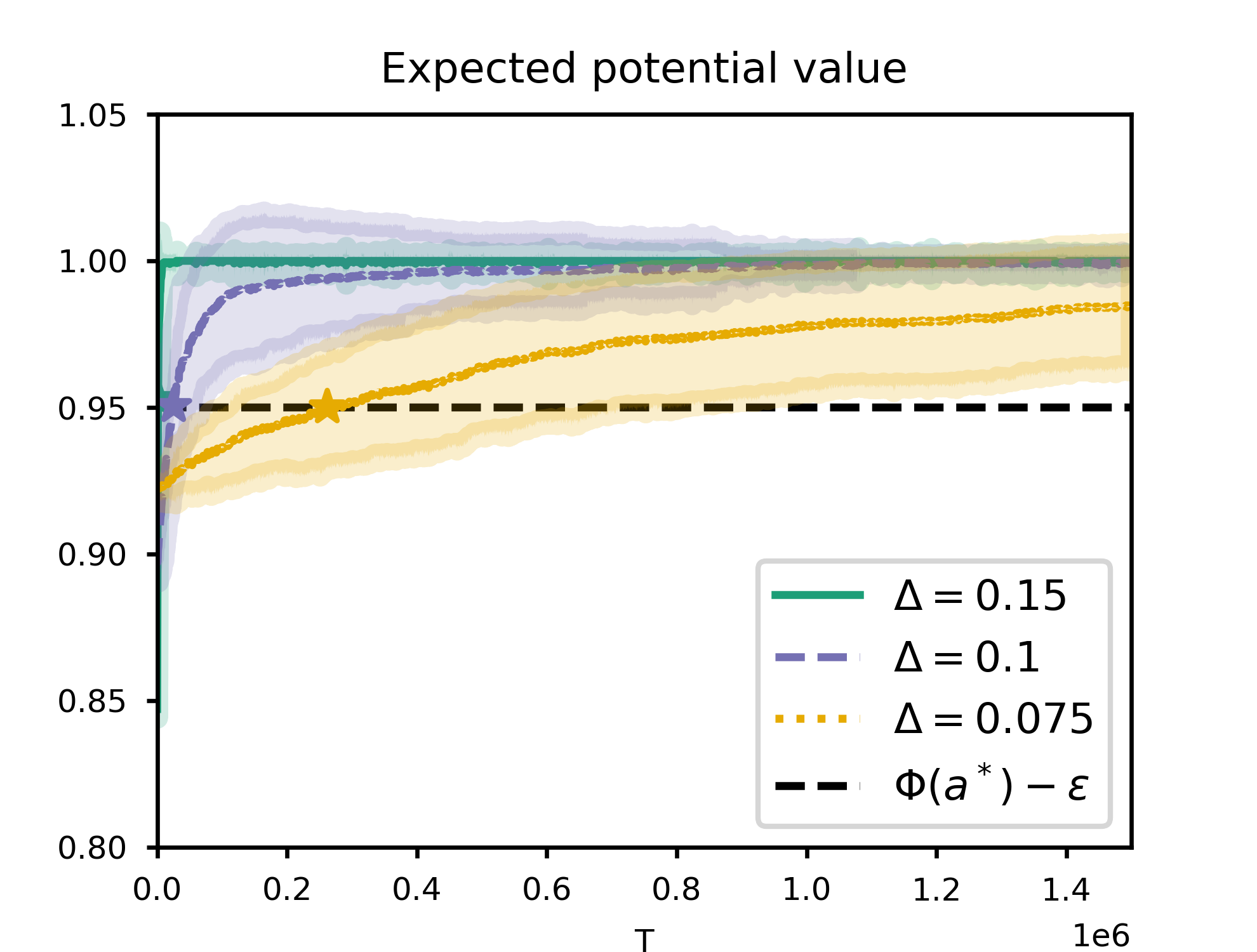}
\end{subfigure}
\begin{subfigure}[b]{0.31\textwidth}
   \includegraphics[width=1\linewidth]{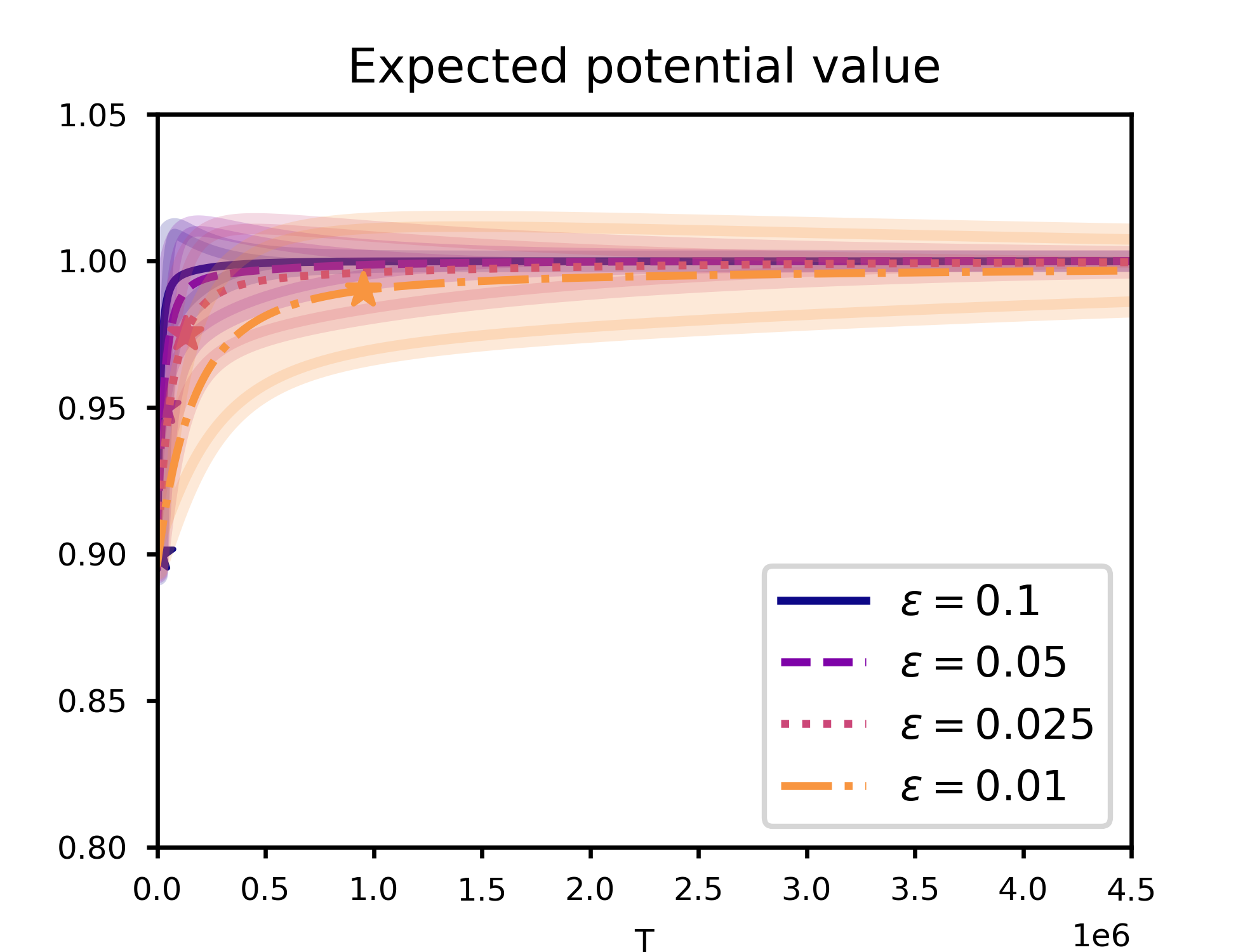}
   \caption{Log-linear learning}
   \label{fig:lll}
\end{subfigure}
\begin{subfigure}[b]{0.31\textwidth}
   \includegraphics[width=1\linewidth]{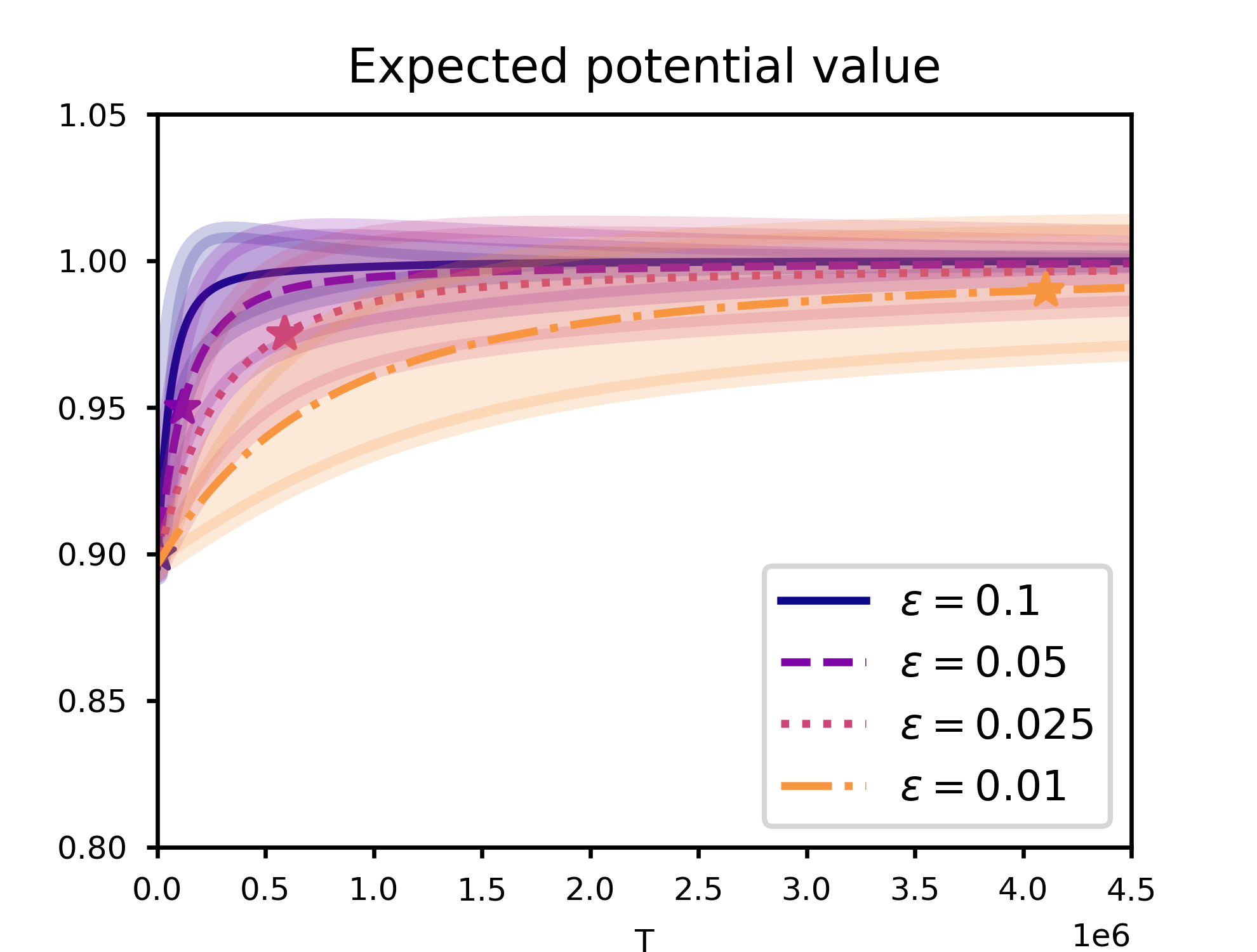}
   \caption{Binary log-linear learning}
   \label{fig:binarylll}
\end{subfigure}
\begin{subfigure}[b]{0.31\textwidth}
   \includegraphics[width=1\linewidth]{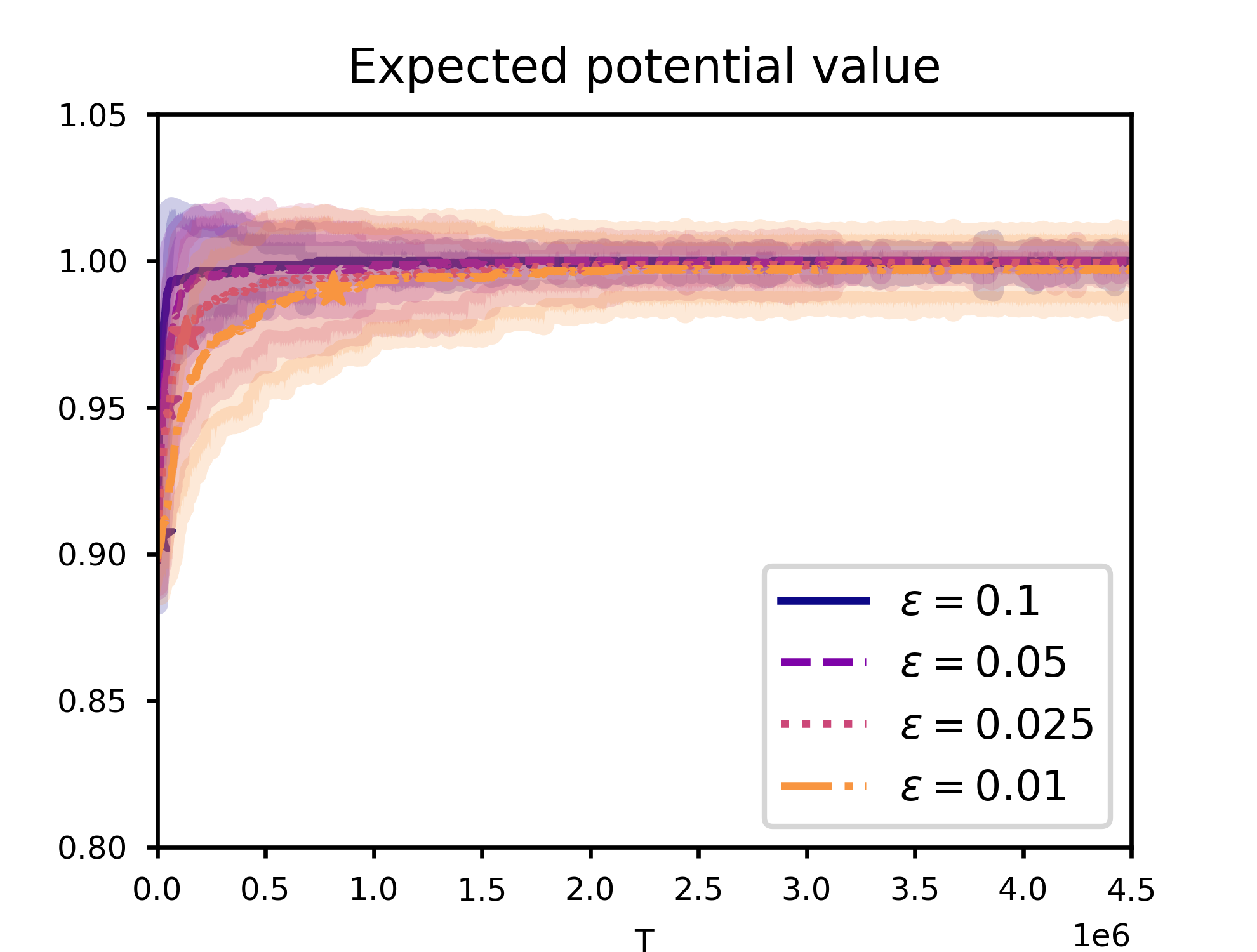}
   \caption{Perturbed log-linear learning}
   \label{fig:perturbedlll}
\end{subfigure}

\caption{Expected potential value when all players follow log-linear learning with $\beta$ set as the lower-bound of Inequality \eqref{eq:beta}. Lines are averages over 30 randomly generated games, shaded areas represent one standard deviation, and stars mark the first time the desired precision $1-\epsilon$ is reached. Top row is given for fixed precision $\epsilon = 0.05$ and various suboptimality gaps $\Delta$, and bottom row for fixed suboptimality gap $\Delta=0.1$ and various precisions $\epsilon$. } 

\label{fig:expected_potential_value}
\end{figure*}

\subsubsection{Log-linear learning mixed with uniform exploration}

We assume players occasionally explore actions randomly. A modification of log-linear learning based on the fixed-share algorithm \cite{herbster1998tracking} can reflect such a random behavior. In the so-called fixed-share log-linear learning, a player $i$ is randomly chosen and allowed to alter her action. Player $i$ samples her new action according to the following strategy:

\begin{align*}
    \hat{p}_i^t(a_i)=\frac{\xi}{A} + \frac{(1-\xi) e^{\beta U_i(a_i,a_{-i}^{t-1})}}{\sum_{a_i'\in\mathcal{A}}e^{\beta U_i(a_i',a_{-i}^{t-1})}},\quad \forall a_i\in\cA.
\end{align*} 
The exploration parameter $\xi\in(0,1)$ determines how likely a player is to act randomly, where a value of $\xi=1$ corresponds to a uniform action sampling while $\xi=0$ corresponds to log-linear learning. For simplicity, we focus on the full-information case, but fixed-share log-linear learning can easily be adapted to the binary setting. Note that this modification resembles the $\epsilon$-Hedge strategy \cite{heliou2017learning} in the expert advice literature, and under binary feedback, this modification resembles the Epx3.P strategy \cite{auer2002nonstochastic,bubeck2012regret} in the bandit literature. Here, the fixed share $\xi/A$ ensures a lower bound on the exploration.


Without knowing the stationary distribution of this learning rule, we can apply Theorem~\ref{thm:General_Convergence} to deduce the following result.

\begin{corollary}\label{cor:fixed_share}
    Consider the setting of Theorem \ref{thm:General_Convergence}, where all players adhere to fixed-share log-linear learning with $\beta~=~\Tilde{\Omega}\left(\frac{1}{\Delta}\log \frac{A^N}{\epsilon}\right)$. Then, for $\epsilon\in(0,1)$ and initial distribution $\mu^0$ we have: 
    
    \begin{equation*}
        \bE_{a\sim \mu^t}[\Phi(a)] \ge \max_{a\in \cA} \Phi(a) - \epsilon - \frac{L \Ap}{\sqrt{N}} \xi,
    \end{equation*}
    for {\small$t=\mO( N^{3/2} A^{N+3} e^{\beta(N+3)} / (1-\xi)^{N+3} )$} with {\small$L = \mathcal{\Tilde{O}}(N^2 A^{N+5} e^{\log(\Ap /\epsilon)/\Delta} )$}. 
\end{corollary}

\noindent The proof follows from applying Theorem~\ref{thm:General_Convergence} and is provided in Appendix~\ref{app:fixed_share}. Corollary~\ref{cor:fixed_share} guarantees the convergence of fixed-share log-linear learning to an $\epsilon$-efficient Nash equilibrium in time polynomial in $1/\epsilon$ if the exploration parameter $\xi$ is sufficiently small. The key is to show that the transition matrix of fixed-share log-linear learning is close to the transition matrix of the unperturbed learning rule in terms of the $\ell_2$ distance.

\section{Numerical Illustrations}
\label{app:experiments}



We illustrate our convergence time results for log-linear learning on identical interest games.\footnote{We provide the code for our experiment \href{https://github.com/hanacatic/potentialgames}{here}.}
Concretely, we consider a two-player game where each player has the action set $\cA=\{1,2,\ldots, 10\}$, and the players have the same utility matrix denoted by $\{U(a_1,a_2)\}_{a_1,a_2 \in [10]}$. We generate $30$ different utility matrices $U(\cdot,\cdot)\in [0,1]^{10\times 10}$ of the following form: we first fix a suboptimality gap $\Delta$, then we set $U(2,2) = 1$, $U(9,9) = 1-\Delta$, referred to as plateaus. We sample the remaining entries of the utility matrix uniformly from the range $[0, 1-\Delta)$, such that they form regions of lower value compared to the two plateaus $U(2,2)$ and $U(9,9)$. 
%



In our first experiment, we fix $\epsilon$ as $0.05$ and vary the suboptimality gap $\Delta$ in $[0.15,0.10,0.075]$. We set the temperature parameter $\beta$ as the lower bound of Inequality \ref{eq:beta}. Figure~\ref{fig:lll} (top) demonstrates that the convergence time increases as the suboptimality gap $\Delta$ decreases.
This shows that the suboptimality gap $\Delta$ quantifies the convergence time of log-linear learning well and therefore should be taken into account when bounding the convergence time, as done in our Theorem~\ref{thm:convergence_rates}. 


In our second experiment, we fix $\Delta$ as $0.1$ and vary the precision $\epsilon$ in $[0.1,0.05,0.025,0.01]$. Figure \ref{fig:lll} (bottom) shows that the convergence time increases as $\epsilon$ decreases. In other words, the convergence time of log-linear learning depends inversely on the precision, which is also reflected in our convergence time of $\Tilde{\mathcal{O}}((\frac{1}{\epsilon})^{1/\Delta})$ in Theorem \ref{thm:convergence_rates}.

The experiments are repeated for binary log-linear learning and perturbed log-linear learning with corrupted utilities. We set $\beta$ as the lower bound on the log-linear learning temperature parameter according to Theorem \ref{thm:convergence_rates_binary} and Corollary \ref{cor:noise_corruption}, respectively.
Figure \ref{fig:binarylll} illustrates that two-point feedback leads to an increase in convergence time, however, the order of the convergence time is the same as for classical log-linear learning. This is consistent with the feedback reduction increasing the convergence time by a constant factor in Theorem \ref{thm:convergence_rates_binary}. Lastly, Figure \ref{fig:perturbedlll} shows that corrupting the utilities with a small bounded noise has a negligible effect on the convergence time of log-linear learning, as proven in Corollary \ref{cor:noise_corruption}.

\begin{figure}[!h]
\begin{subfigure}[b]{0.48\textwidth}
   \includegraphics[width=1\linewidth]
   {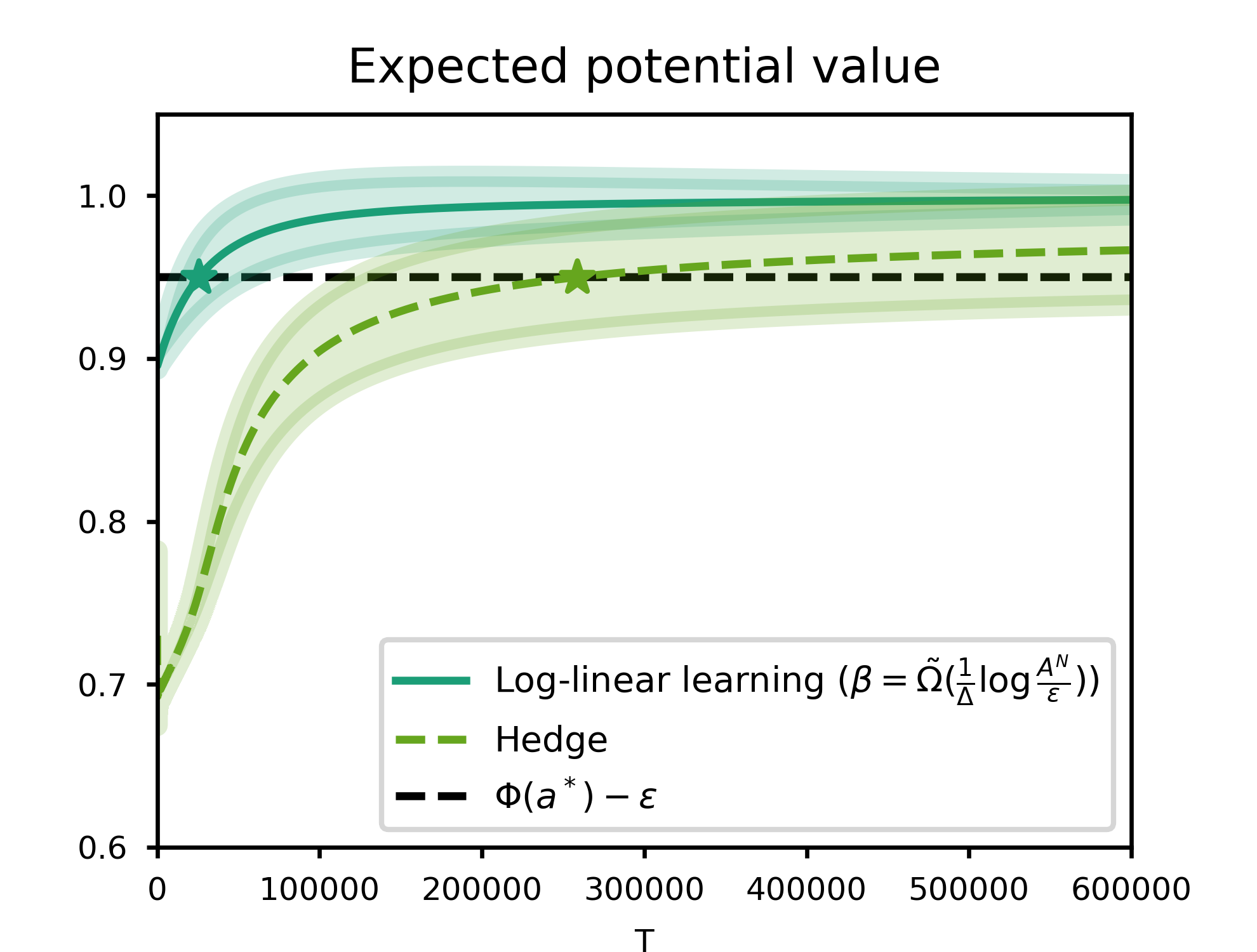}
   \caption{Full feedback}
\end{subfigure}
\begin{subfigure}[b]{0.48\textwidth}
   \includegraphics[width=1\linewidth]{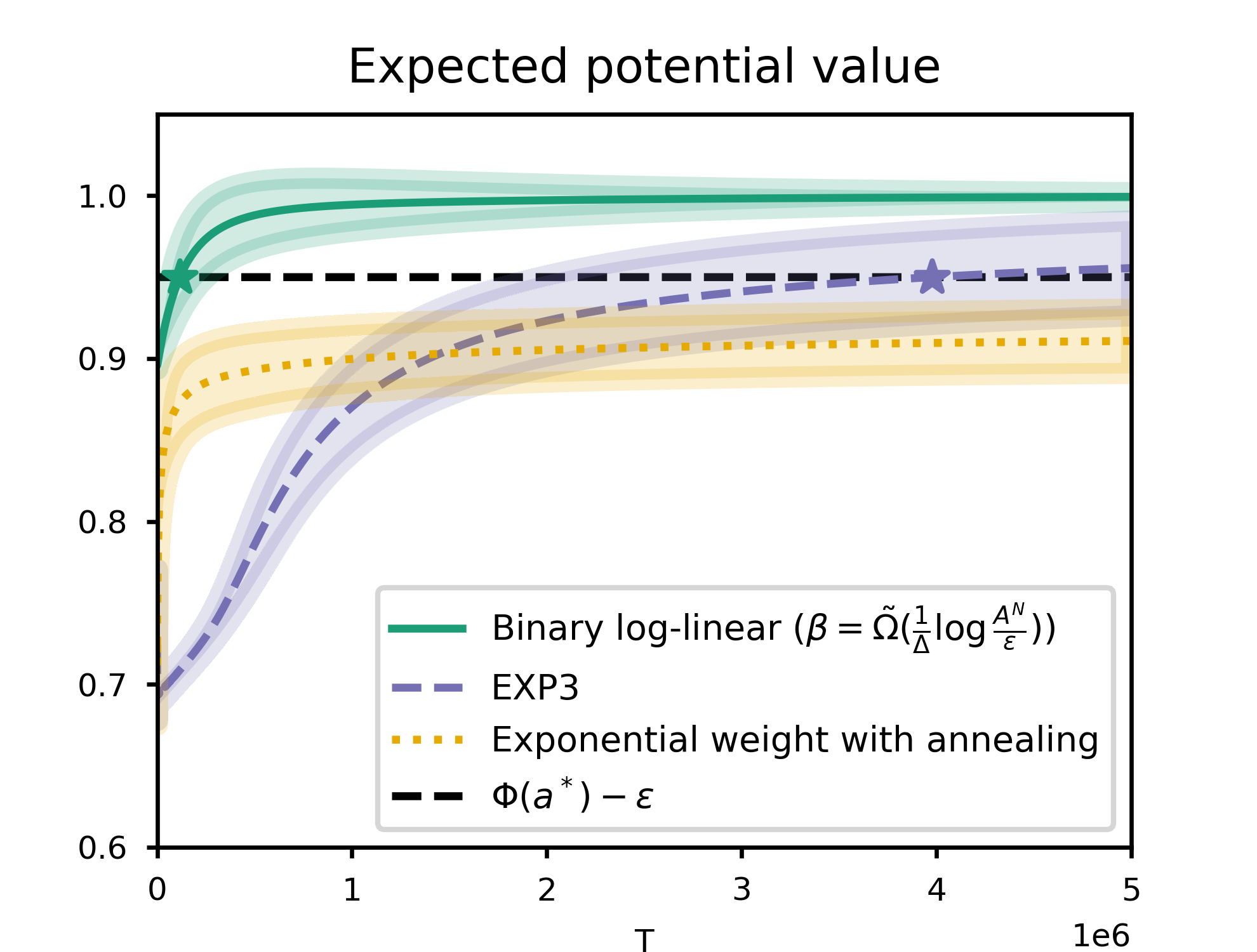}
   \caption{Reduced feedback}
   \label{fig:comparison_b}
\end{subfigure}

\caption{Comparison of log-linear learning. Lines are averages over 30 randomly generated games, shaded areas represent one standard deviation, and stars mark the first time the desired precision $1-\epsilon$ is reached. The results are given for a fixed precision $\epsilon=0.05$}
\label{fig:comparison}
\end{figure}

Finally, we compare log-linear learning with Hedge \cite{freund1997decision}, both of which rely on full-information feedback, as well as binary log-linear learning with the exponential weights algorithm for exploration and exploitation (EXP3) \cite{auer2002nonstochastic} and the exponential weights algorithm with annealing \cite{heliou2017learning}, which operate under reduced feedback. In each case, we set $\beta$ according to Theorems \ref{thm:convergence_rates} and \ref{thm:convergence_rates_binary}, respectively. Figure \ref{fig:comparison} shows that both log-linear learning and binary log-linear learning achieve convergence to an $\epsilon$-efficient Nash equilibrium with faster convergence times. Notably, Figure \ref{fig:comparison_b} illustrates that the exponential weights algorithm with annealing fails to converge to an $\epsilon$-efficient Nash equilibrium. This observation is consistent with theoretical results, which guarantee convergence only to a Nash equilibrium for Hedge, EXP3, and exponential weights with annealing.

\section{Conclusion}

We provided the first finite-time convergence guarantees to an $\epsilon$-efficient Nash equilibrium for potential games using a novel mixing-time bound based on a log-Sobolev constant. In particular, using a problem-dependent analysis, we guarantee a polynomial dependence on $1/\epsilon$ for constant $\epsilon>0$. Furthermore, under the additional assumption that the game is symmetric, we showed that the exponential dependence on the number of players $N$ present in our bound can be avoided.  
To deal with reduced feedback, we considered binary log-linear learning and showed that it enjoys the same convergence time as log-linear learning up to numerical constants. 
We also proved that the convergence time of log-linear is not hindered by corruptions of the utilities by bounded noise or by small perturbations in the learning rule. Lastly, we validated our results in a numerical case
study on identical interest games. 



\bibliographystyle{plainnat}
\bibliography{Bibliography}
\appendix

\section{Background on Markov chains and mixing times}\label{app:markov_chains}

Consider a time-homogeneous Markov chain $\{X_t\}_{t\in\N}$ over the state space $\cAp$ with transition matrix $P\in\R^{{\Ap}\times \Ap}$. The ergodic theorem \cite{levin2017markov} ensures that an irreducible and aperiodic Markov chain $\{X_t\}_{t\in\N}$ has a unique stationary distribution $\mu$, and from any initial distribution $\mu^0$ the distribution $\mu^t=\mu^0P^t$ converges to $\mu$. The convergence time to the stationary distribution is quantified by the mixing time:

\begin{align}\label{eq:mixing_time}
    \tm{P}(\epsilon) \eqdef \min\{t\in\bN \: | \: \|\mu^t-\mu\|_{TV}\leq \epsilon\},
\end{align}
where the total variation distance is defined as $\|\mu^t-\mu\|_{TV} \eqdef \frac{1}{2}\sum_{a\in\cAp}|\mu^t(a)-\mu(a)|$ \cite{montenegro2006mathematical}. Next, we provide a bound on the mixing time of $\{X_t\}_{t\in\N}$ based on the log-Sobolev constant.

\begin{lemma}{\cite[Section~3]{diaconis1996logarithmic}}\label{lem:general_mixing_time}
    If $P$ is irreducible and aperiodic, then the mixing time has the following upper bound: 
    \begin{align}\label{eq:mixing_time_bound}
    \tm{P}(\epsilon) \le \frac{1}{\rho(PP^*)}\left(\log\log \frac{1}{\mu_{\min}}+2\log\frac{1}{\epsilon}\right),
\end{align}
    where $\mu_{\min}\eqdef \min_{a\in\Ap}\mu(a)$, $P^*$ is the time-reversal of $P$, and $\rho(PP^*)$ denotes the log-Sobolev constant of $PP^*$ defined as:\footnote{$P^*$ satisfies $\mu(a)P^*(a,\Ta)=\mu(\Ta)P(\Ta,a) \forall a,\Ta \in \cAp$. The chain is called time-reversible if $P^*=P$.}
    
    \begin{align}\label{eq:def_log_Sobolev}
        \rho(P)\eqdef\inf_{\mL_\pi (f^2)\neq 0} \frac{\mE_P(f,f)}{\mL_\pi (f^2)},
    \end{align}
    where for $f:\cAp\rightarrow\R$, the Dirichlet form is defined as:

$$\mE_P(f,f)=\langle f,(I-P)f\rangle_\pi=\frac{1}{2}\sum_{a,\Ta\in\Ap}(f(a)-f(\Ta))^2P_{a,\Ta}\mu(a),$$
and the entropy-like quantity $\mL(f^2)$ is given by:

$$\mL(f^2)=\sum_{a\in\Ap} f(a)^2\log\frac{f(a)^2}{\|f\|_2^2}\mu(a).$$

\end{lemma}

\noindent We briefly compare this mixing time bound to classical ones based on the spectral gap $\lambda(P)$, which are of the form \cite{montenegro2006mathematical}: 

\begin{align*}
    \tm{P}(\epsilon) \le \frac{C}{\lambda(PP^*)}\bigg(\log \frac{1}{\sqrt{\mu_{\min}}}+\log \frac{1}{\epsilon}\bigg),
\end{align*}
where $C$ is a constant and the spectral gap $\lambda(P)$ is:

    \begin{align}\label{eq:def_spectral_gap}
        &\lambda(P)\eqdef \inf_{\text{Var}_\pi (f)\neq 0} \frac{\mE_P(f,f)}{\text{Var}_\pi (f)},
    \end{align}
    where $\text{Var}_\pi(f)=\sum_{a,\Ta\in\Ap} (f(a)-f(\Ta))^2\mu(a)\mu(\Ta)$.
    
    \noindent Mixing time bounds using log-Sobolev constants are often significantly tighter than those based on the spectral gap. 
    Indeed, in Lemma $3.1$ of \cite{diaconis1996logarithmic} it is shown that the log-Sobolev constant $\rho(PP^*)$ is upper-bounded by the spectral gap $\lambda(PP^*)$ as follows: $2\rho(PP^*)\le\lambda(PP^*)$. Thus, if 

\begin{align}\label{eq:spectral-gap-comparison}
    \log\log \frac{1}{\mu_{\min}}\le \log \frac{1}{\sqrt{\mu_{\min}}},
\end{align}
then, the mixing time bound based on the log-Sobolev constant improves over the spectral gap counterpart. To illustrate, consider a Markov chain on the $d$-dimensional hypercube $\mathcal{H}=\{-1,1\}^d$ with uniform stationary distribution. Then, $\mu_{\min}=2^{-d}$ and Equation \eqref{eq:spectral-gap-comparison} is satisfied in this example. However, deriving log-Sobolev constants can be extremely difficult, and thus, the corresponding bounds are less explored.

\hide{
\begin{proof}(Lemma \ref{lem:general_mixing_time})\label{app:general_mixing_time}
    Let the relative entropy be defined as {\small$D(\mu^t:\mu) \eqdef \sum_{a\in\cAp} \mu^t(a)\log \frac{\mu^t(a)}{\mu(a)}$}. Then, for a Markov chain $\{X_t\}_{t\in\N}$ with irreducible transition matrix $P$ the relative entropy $D(\mu^t:\mu)$ decays at the following rate \cite{miclo1997remarques}:

\begin{align}\label{eq:sobolev_bound}
    D(\mu^t:\mu)\leq(1-\rho(PP^*))^t D(\mu^0:\mu).
\end{align}
Using Pinsker's inequality we have that:

\begin{align}\label{eq:rel_entropy}
    \|\mu^t-\mu\|_{TV}\leq \sqrt{\frac{D(\mu^t:\mu)}{2}}\le\sqrt{(1-\rho(PP^*))^t D(\mu^0:\mu)}.
\end{align}

    Note that $\rho(PP^*)<1$ since $2\rho(PP^*)\le \lambda(PP^*)$ by Lemma 3.1 in \cite{diaconis1996logarithmic} and for the spectral gap $\lambda(PP^*)$ it is known that $\lambda(PP^*)<1$ \cite{levin2017markov}. To ensure that $\|\mu^t-\mu\|_{TV}\le\epsilon$, we derive the following lower bound on $t$:
 
    \begin{align*}
        & \sqrt{(1-\rho(PP^*))^t D(\mu^0:\mu)}\le \epsilon\\
        \Leftrightarrow\quad & t\underbrace{\log(1-\rho(PP^*))}_{<0}\le\log\bigg(\frac{\epsilon^2}{D(\mu^0:\mu)}\bigg)\\
        \Leftrightarrow\quad & t\ge \frac{1}{\log(1-\rho(PP^*))}\log\bigg(\frac{\epsilon^2}{D(\mu^0:\mu)}\bigg)\\
        \Leftrightarrow\quad & t\ge -\frac{1}{\rho(PP^*)}\log\bigg(\frac{\epsilon^2}{D(\mu^0:\mu)}\bigg)\\
        \Leftrightarrow\quad & t\ge \frac{1}{\rho(PP^*)}\bigg(\log(D(\mu^0:\mu)+2\log\bigg(\frac{1}{\epsilon}\bigg)\bigg)\\
        \Leftrightarrow\quad & t\ge \frac{1}{\rho(PP^*)}\bigg(\log\log\bigg(\frac{1}{\mu_{\min}}\bigg)+2\log\bigg(\frac{1}{\epsilon}\bigg)\bigg),
    \end{align*}
    where in line $4$ we used that $\log(1+x)\le x$ for $x>-1$ and then in line $6$  we used that for any $\mu_0$, $D(\mu^0:\mu)\leq \log \frac{1}{\mu_{\min}}$ for $\mu_{\min} \eqdef \min_{a\in\cAp}\mu(a)$.
    Thus, we conclude that the mixing time is upper-bounded as follows:
    
\begin{align}
    \tm{P}(\epsilon) \le \frac{1}{\rho(PP^*)}\bigg(\log\log \frac{1}{\mu_{\min}}+2\log\frac{1}{\epsilon}\bigg).
\end{align} 

\end{proof}
}

\section{Convergence of log-linear learning}\label{app:thm1}

Here, we provide a proof of Lemma \ref{lem:beta_bound} and Corollary \ref{cor:symmetric_potential}.

\subsection{Proof of Lemma \ref{lem:beta_bound}}\label{app:proof_beta_bound}
    Define the set $\cAp_*=\{a^*\in\cAp\: | \:a^*\in\argmax_{a\in\cAp}\!\!\Phi(a)\}$ 
    as the set of potential maximizers with cardinality $\!\Ap_* \!=\!|\cAp_*|$. Then, the expected value of the potential function $\Phi(\cdot)$ over the stationary distribution $\mu$ of log-linear learning in \eqref{eq:stationary_dist_log_linear} can be bounded as follows:

    \begin{align}
            &\mathbb{E}_{a\sim\mu}[\Phi(a)] = \sum_{a\in\cAp}\frac{e^{\beta\Phi(a)}}{\sum_{\Tilde{a}\in\cAp}e^{\beta\Phi(\Tilde{a})}}\Phi(a) \nonumber\\
            &\geq \sum_{a\in\cAp_*}\frac{e^{\beta\Phi(a)}}{\sum_{\Tilde{a}\in\cAp}e^{\beta\Phi(\Tilde{a})}}\Phi(a) \nonumber\\
            &=\sum_{a\in\cAp_*}\frac{\Phi(a)}{\sum_{\Tilde{a}\in\cAp_*}e^{\beta(\Phi(\Tilde{a})-\Phi(a))}+\sum_{\Tilde{a}\in\cAp\setminus\cAp_*}e^{\beta(\Phi(\Tilde{a})-\Phi(a))}} \nonumber\\
            &\ge\sum_{a\in\cAp_*}\frac{\Phi(a)}{\sum_{\Tilde{a}\in\cAp_*}e^{0}+\sum_{\Tilde{a}\in\cAp\setminus\cAp_*}e^{-\beta\Delta}} \nonumber\\
            &\geq \frac{\Ap_*}{\Ap_* + (\Ap - \Ap_*)e^{-\beta\Delta}}\Phi(a^*). \label{eq:lower_bound_potential}
        \end{align}
    where the suboptimality gap $\Delta$ is given by $\Delta \eqdef \min_{a\in\cAp: \Phi(a) < \Phi(a^*)}\left(\Phi(a^*)-\Phi(a)\right)$ with $a^*\in\cAp_*$. If
    \begin{align}\label{eq:value_beta}
        \beta\ge\frac{1}{\Delta}\log((A^N-A^N_*)(\frac{1}{\epsilon A_*^N} - \frac{1}{A_*^N})),
    \end{align} 
    then $\frac{\Ap_*}{\Ap_* + (\Ap - \Ap_*)e^{-\beta\Delta}}\ge 1-\epsilon$. If $\beta$ is set as in Equation \eqref{eq:value_beta}, injecting the last inequality into Equation \eqref{eq:lower_bound_potential} implies:
    \begin{align*}
        \mathbb{E}_{a\sim\mu}[\Phi(a)]\geq (1-\epsilon)\Phi(a^*) = \Phi(a^*)-\epsilon,
    \end{align*}
    where we used that $\Phi(a^*)\le 1$. This concludes the proof.

\hide{
\subsection{Proof of Lemma \ref{lem:general_bound_log_Sobolev}}\label{app:proof_general_bound_log_Sobolev}

We first state the following lemma which will be used to prove Lemma \ref{lem:general_bound_log_Sobolev}.
\begin{lemma}\label{lem:helper_log-Sobolev_bound}
    Consider the Markov chain $\{\hX_t\}_{t\in\N}$ over a finite state space $\cAp$ with transition matrix $\hP\in\R^{\Ap\times\Ap}$ given by:
    
    \begin{align}\label{eq:helper_log-Sobolev_bound}
    \hP_{a,\Ta}=
    \frac{1}{NA}\mathds{1}_{\Ta \in \cN(a)} 
\end{align}
where $\cN(a) = \{\Ta\in\cAp \: | \:\exists i\in [N]: \Ta_{-i} = a_{-i}\}$.
Assuming $A\ge 4$, the log-Soblev constant of $\hP$ is lower-bounded by:

\begin{align*}
        \rho(\hP)\geq \frac{4\pi^2}{25NA}.
    \end{align*}
\end{lemma}
\ \\
\begin{proof}
    We will lower bound the log-Soblev constant of $\hP$ in terms of the log-Soblev constant of another Markov chain for which a lower bound on the log-Soblev constant is known.

    First, note that the Markov chain $\{\hX_t\}_{t\in\N}$ is aperiodic and irreducible with stationary distribution $\hmu(a)=1/\Ap$. This can be verified by checking the detailed balance equations given by $\hmu(a)\hP_{a,\Ta}=\hmu(\Ta)\hP_{\Ta,a}$ for all $a,\Ta\in\Ap$.  Next, we make use of Corollary $2.15$ in \cite{montenegro2006mathematical} to lower-bound $\rho(\hP)$ in terms of the log-Soblev constant of another Markov chain $\bX_t$ with transition matrix $\bP$ and stationary distribution $\bmu$. Namely, 
    
    \begin{align*}
        \rho(\hP)\geq \frac{1}{MC}\rho(\bP),
    \end{align*}
    where
    
    \begin{align*}
        &M=\max_{a\in\cAp}\frac{\hmu(a)}{\bmu(a)},\\
        & C = \max_{a\neq\Ta: \hP_{a,\Ta}\neq 0} \frac{\bmu(a)\bP_{a,\Ta}}{\hmu(a)\hP_{a,\Ta}}.
    \end{align*}
    As the comparison Markov chain, we consider the product chain $\{\bX_t\}_{t\in\N}$ with $\bX_t= \prod_{i=1}^N\bX_{i,t}$ on the state space $\mathbb{Z}_{\Kp}=\prod_{i=1}^N\mathbb{Z}_{K}$. Here, each $\{\bX_{i_t}\}_{t\in\N}$ is a simple random walk on the finite circle $\Z_{K}=\{1,\ldots,K\}$ with $K\geq 4$. The transition matrix and the stationary distribution of the simple random walk $\bX_{i_t}$ are given by $\bP_{i_{k,k\pm 1}}=\frac{1}{2}$ and $\bmu_i(k)=1/K$, respectively. Furthermore, the log-Soblev constant $\rho(\bP_i)$ is lower bounded by
        $\rho(\bP_i)\geq \frac{8\pi^2}{25 K^2}$ \cite[Example~4.2]{diaconis1996logarithmic}.
  Thus, by definition the product chain $\bX_t$ \cite[Sec.~2.5]{diaconis1996logarithmic} has the following transition matrix:
  
    \begin{align*}
        \bP_{\textbf{k},\Tilde{\textbf{k}}}=
\frac{1}{2N}\mathds{1}_{\Tilde{\textbf{k}}=(k_i\pm 1,\textbf{k}_{-i})},
    \end{align*}
   and stationary distribution:

\begin{align*}
    \bmu(\textbf{k})=\prod_{i=1}^N \bmu_i(k_i) =\prod_{i=1}^N \frac{1}{K}=\frac{1}{\Kp}.
\end{align*}
Furthermore, by Lemma $3.2$ in \cite{diaconis1996logarithmic} the log-Soblev constant $\rho(\bP)$ of the product chain $\{\bX_t\}_{t\in\N}$ is lower bounded by:

\begin{align}\label{eq:log-Sobolev_simple_walk}
    \rho(\bP)=\frac{1}{N}\min_{i\in\{1,\ldots,N\}}\rho(\bP_i)\geq   \frac{8\pi^2}{25NK^2}.
\end{align}
Next, note that there is a one-to-one mapping between the set $\Ap$ and the set $\mathbb{Z}_{K}$ with $|\cA|=A=K$ and thus a one-to-one mapping between the set $\cAp$ and the set $\mathbb{Z}_{\Kp}$ with  with $|\cAp|=\Ap=K^N$. Therefore, we can assume that the Markov chains $\hX_t$ and $\bX_t$ operate on the same state space. By Equation \eqref{eq:log-Sobolev_simple_walk} and with the following upper bounds on $M$ and $C$:

\begin{align*}
    &M=\max_{a\in\cAp}\frac{\hmu(a)}{\bmu(a)}=\frac{\Ap}{\Ap}=1\\
    &C = \max_{a\neq\Tilde{a}: \hP_{a,\Tilde{a}}\neq 0} \frac{\bmu(a)\bP_{a,\Tilde{a}}}{\hmu(a)\hP_{a,\Tilde{a}}}=\frac{A}{2}.
    \end{align*}
the log-Soblev constant $\rho(\hP)$ can be lower-bounded by:

\begin{align*}
        \rho(\hP)\geq \frac{1}{MC}\rho(\bP)\geq \frac{16\pi^2}{25NA^3},
    \end{align*}
    which concludes the proof.
\end{proof}
\noindent
Next, we proceed to prove Lemma \ref{lem:general_bound_log_Sobolev}.
\\
\begin{proof}(Lemma \ref{lem:general_bound_log_Sobolev})
The Markov chain $X_t$ with transition matrix $P$ defined in \eqref{eq:general_form} is aperiodic and irreducible and thus a unique stationary distribution $\mu$ exists with $\mu^t=\mu^0P^t\rightarrow\mu$ for $t\rightarrow\infty$ from any initial distribution $\mu^0$. We define $\mu_{\max}$ and $\mu_{\min}$ as $\max_{a\in\mA}\mu(a)\le\mu_{\max}\leq 1$ and $\min_{a\in\mA}\mu(a)\ge\mu_{\min}>0$, where $\mu_{\min}>0$ follows from the irreducibility of $X_t$.

Next, we consider the modified Markov chain $\{X^*_t\}_{t\in\N}$ with transition matrix $PP^*$ which is aperiodic and irreducible since $X_t$ is aperiodic and irreducible. Concretely, since $P$ contains self-loops, \ie $P_{a,a}>0$, it follows that $PP^*$ contains self-loops, \ie

\begin{align*}
    PP^*_{a,a}&= \sum_{a'\in\mA}P_{a,a'}P^*_{a',a}= \sum_{a'\in\mA}P_{a,a'}\frac{\mu(a)P_{a,a'}}{\mu(a')}\\
    &\geq P_{a,a}P_{a,a}>0,
\end{align*}
and thus $X^*_t$ is aperiodic. Furthermore, for any $a,\Ta\in\cAp$:

\begin{align*}
    (PP^*)^N_{a,\Ta}&=\sum_{\substack{a_l\in\cAp\\l=1,\ldots,N-1}}(PP^*)_{a,a_1}\ldots(PP^*)_{a_{N-1},\Ta}\\
    &=\sum_{\substack{a_l\in\cAp\\l=1,\ldots,N-1}}\sum_{a'\in\cAp}P_{a,a'}P^*_{a',a_1}\ldots\sum_{a'\in\cAp}P_{a_{N-1},a'}P^*_{a',\Ta}\\
    &\geq\sum_{\substack{a_l\in\cAp\\l=1,\ldots,N-1}}P_{a,a_1}P_{a_1,a_1}\ldots P_{a_{N-1},\Ta}P_{\Ta,\Ta}>0,
\end{align*}
where we used that $P^N_{a,\Ta}>0$ and $P_{a,a}>0$ for all $a,\Ta\in\cAp$ as well as the identity $\mu(a)P^*_{a,\Ta}=\mu(\Ta)P_{\Ta,a}$. It follows that $X^*_t$ is irreducible. Thus, for $X^*_t$ a unique stationary distribution $\mu^*$ exists. By Proposition $1.23$ in \cite{levin2017markov} the stationary distribution of $P^*$ is given by $\mu$ and thus the stationary distribution of $PP^*$ is given by $\mu$ since $\mu PP^*=\mu P^*=\mu$. Furthermore, the following holds for the transition matrix $PP^*$ :

\begin{align*}
  \frac{1}{N}p_{\min}^2\mathds{1}_{\Ta \in \cAp(a)}
\leq (PP^*)_{a,\Ta}
    \le\mathds{1}_{\Ta \in \cAp(a)}
\end{align*}
where 

\begin{align*}
    (PP^*)_{a,\Ta}=\sum_{a'\in\mA}P_{a,a'}P^*_{a',\Ta}\geq P_{a,\Ta}P^*_{\Ta,\Ta}\geq P_{a,\Ta}P^*_{\Ta,\Ta}\geq \frac{p_{\min}^2}{N}.
\end{align*}
Next, we compute $M$ and $C$, defined bellow, of the Markov chains $X^*_t$ and $X_t$:

    \begin{align*}
    &M=\max_{a\in\cAp}\frac{\mu(a)}{\mu(a)}=1\\
    &C =\max_{a\neq\Tilde{a}: (PP^*)_{a,\Tilde{a}}\neq 0} \frac{\mu(a)P_{a,\Tilde{a}}}{\mu(a)(PP^*)_{a,\Tilde{a}}}\leq  \frac{N}{ p_{\min}^2}.
    \end{align*}
From Corollary $2.15$ in \cite{montenegro2006mathematical}, it follows that the log-Soblev constant $\rho(PP^*)$ is lower-bounded by:
    
    {\small
    \begin{align}\label{eq:term1_general}
        \rho(PP^*)\geq\frac{1}{MC}\rho(P)\geq \frac{ p_{\min}^2}{N} \rho(P).
    \end{align}
    }
    
Next, we compare the Markov chain $X_t$ to the Markov chain $\hX_t$ with transition matrix $\hP$ specified in Equation \eqref{eq:helper_log-Sobolev_bound} of Lemma \ref{lem:helper_log-Sobolev_bound}. To this end, we compute $M$ and $C$ of the Markov chains $X_t$ and $\hX_t$:

{\small
    \begin{align*}
        &M=\max_{a\in\cAp}\frac{\mu(a)}{\hmu(a)}\leq \Ap\\
    &C = \max_{a\neq\Tilde{a}: P_{a,\Tilde{a}}\neq 0} \frac{\hmu(a)\hP_{a,\Tilde{a}}}{\mu(a)P_{a,\Tilde{a}}}
    \leq  \frac{N}{\Ap NA\mu_{\min}p_{\min}},
    \end{align*}
    }
    
Thus, by Corollary $2.15$ in \cite{montenegro2006mathematical} and by Lemma \ref{lem:helper_log-Sobolev_bound} the log-Soblev constant $\rho(P)$ can be lower-bounded by:

{\small
\begin{align}
        \rho(P)\geq \frac{1}{MC}\rho(\hat{P})\geq \Ap A\mu_{\min}p_{\min}\rho(\hat{P})\geq \frac{16\pi^2\Ap\mu_{\min}p_{\min}}{25NA^2}.
    \end{align}
    }
    
Combining Equation \eqref{eq:term1_general} and \eqref{eq:term2_general}, we conclude that the log-Soblev constant $\rho(PP^*)$ is lower-bounded by:

{\small
    \begin{align*}
        \rho(PP^*)\geq \frac{16\pi^2\Ap\mu_{\min} p_{\min}^3}{25N^2A^2}.
    \end{align*}
    }
    

    

\end{proof}
}

\subsection{Proof of Corollary \ref{cor:symmetric_potential}}\label{app:modified_log_linear}

    Define the set $\Psi_*^\cA=\{x^*\in\Psi^\cA\: | \:x^*\in\argmax_{x\in\Psi^\cA}\!\!\Phi_m(x)\}$ 
    as the set of potential maximizers with cardinality $\!Y_* \!=\!|\Psi_*^\cA|$. Then, the expected value of the potential function $\Phi_m$ over the stationary distribution $\mu_m$ of modified log-linear learning in \eqref{eq:stationary_dist_log_linear} can be bounded as follows:
{\small
    \begin{align}
            &\mathbb{E}_{x\sim\mu}[\Phi_m(x)] = \sum_{x\in\Psi^\cA}\frac{e^{\beta\Phi_m}(x)}{\sum_{\Tx\in\Psi^\cA}e^{\beta\Phi_m}(\Tx)}\Phi_m(x) \\
            &\geq\!\!\!\!\sum_{x\in\Psi_*^\cA} \!\! \frac{\Phi_m}(x){\sum_{\Tx\in\Psi_*^{\mc A}}e^{\beta(\Phi_m(\Tx)-\Phi_m(x))}\!+\!\sum_{\Tx\in\Psi^\cA\!\setminus\!\Psi_*^{\mc A}}e^{\!\beta(\Phi_m(\Tx)-\Phi_m(x))}}\nonumber\\
            &\ge\sum_{a\in\Psi_*^\cA}\frac{\Phi_m(x)}{\sum_{\Tx\in\Psi_*^{\mc A}}e^{0}+\sum_{\Tx\in\Psi^\cA\setminus\Psi_*^{\mc A}}e^{-\beta\Delta}} \nonumber\\
            &\geq \frac{Y_*}{Y_* + (Y - Y_*)e^{-\beta\Delta}}\Phi_m(x^*), \nonumber
        \end{align}
        }   
        
    \noindent
    where $\Delta \eqdef \min_{x\in\Psi^\cA: \Phi_m(x) < \Phi_m}(x^*)\left(\Phi_m(x^*)-\Phi_m(x)\right)$ is the suboptimality gap with $x^*\in\Psi_*^\cA$. Then, for
    {\small
    \begin{align}\label{eq:beta_modified_log_linear}
        \beta \ge 
        &\frac{1}{\Delta}\log\bigg((N+1)^{A-1}\bigg(\frac{1}{\epsilon Y_*}-\frac{1}{Y_*}\bigg)\bigg),  
    \end{align}
    }
    \noindent It holds that: 
    {\small
    \begin{align*}
        \frac{Y_*}{Y_* + (Y - Y_*)e^{-\beta\Delta}}\geq 1-\epsilon,
    \end{align*}
    }
    
    \noindent where we used that $Y\le (N+1)^{A-1}$.
   We deduce that for {\small$\beta=\Omega\left(\frac{1}{\Delta}\log\left(\frac{N^A}{\epsilon}\right)\right)$}, it holds that:
   {\small
    \begin{align*}
        \mathbb{E}_{x\sim\mu_m}[\Phi_m(x)] \geq (1-\epsilon)\max_{x\in\Psi^\cA}\Phi_m(x) \geq \max_{x\in\Psi^\cA}\Phi_m(x)-\epsilon.
    \end{align*}
    }
    
    \noindent The proof now follows from the same analysis as in the proof of Theorem $3$ in \cite{shah2010dynamics} with the exception that we replace Lemma $6$ in \cite{shah2010dynamics} with our analysis above. Concretely, we set $\beta$ as specified in Equation \eqref{eq:beta_modified_log_linear} rather than as in \cite[Eq.~(8)]{shah2010dynamics}.

\section{Binary log-linear learning}\label{app:thm2}

Binary log-linear learning induces an irreducible and aperiodic Markov chain $\{X_t\}_{t\in\Z_+}$ with a time-reversible transition matrix $P\in\R^{\cA\times \cA}$ given by:

\begin{align}\label{eq:transition_prob_payoff_log_linear}
    P_{a,\Ta} = \frac{1}{N}\frac{1}{A}\frac{e^{\beta U_i(\Ta_i,\Ta_{-i})}}{e^{\beta U_i(a_i,\Ta_{-i})} + e^{\beta U_i(\Ta_i,\Ta_{-i})}} \mathds{1}_{\Ta \in \cN(a)} 
\end{align}
where $\cN(a) = \{\Ta\in\cAp \: | \:\exists i\in [N]: \Ta_{-i} = a_{-i}\}$. The additional term $1/A$ stems from the fact that player $i$ first randomly samples an action $\Ta_i$ and then decides between this action and her previous action. \cite{arslan2007autonomous} show that its stationary distribution $\mu\in\Delta(\cAp)$ is given by:

\begin{align}\label{eq:stationary_dist_payoff_log_linear}
    \mu(a) = \frac{e^{\beta\Phi(a)}}{\sum_{\Ta\in\Ap}e^{\beta\Phi(\Ta)}} \quad\forall a\in \cAp.
\end{align}
Importantly, note that the stationary distribution of binary log-linear learning is the same as that of log-linear learning (Equation \eqref{eq:stationary_dist_log_linear}). Thus, log-linear and binary log-linear learning converge to an approximately efficient Nash equilibrium in the long run.
We briefly outline the proof of Theorem \ref{thm:convergence_rates_binary} and then provide a detailed proof. 

\paragraph{Proof outline:} The proof follows from the same line of arguments as in the proof of Theorem \ref{thm:convergence_rates}. In particular, the first step in the proof of Theorem \ref{thm:convergence_rates} remains the same since binary log-linear learning has the same stationary distribution as log-linear learning. Compared to the second step in the proof of Theorem \ref{thm:convergence_rates}, the main difference is that the transition matrix defined in \eqref{eq:transition_prob_payoff_log_linear} of binary log-linear learning differs from the transition matrix defined in \eqref{eq:transition_prob_log_linear} of log-linear learning. Thus, the log-Sobolev constant of binary log-linear can be lower-bounded as follows:

    \begin{align}\label{eq:log-Sobolev_binary}
        \rho(PP^*)\ge \frac{16\pi^2 \Ap\mu_{\min} p_{\min}^3}{25N^2A^2}\geq \frac{2\pi^2 e^{-4\beta}}{25N^2 A^5},
    \end{align}
    while the log-Sobolev constant of log-linear can be lower-bounded as follows:

    \begin{align*}
        \rho(PP^*)\geq \frac{16\pi^2 \Ap\mu_{\min} p_{\min}^3}{25N^2A^2}\geq \frac{16\pi^2 e^{-4\beta}}{25N^2 A^5}.
\end{align*}
Then, we use Lemma \ref{lem:general_mixing_time} to show that 

    $$\|\mu^t-\mu\|_{TV}\le \epsilon/4
    $$ 
    for {\small$t\ge\frac{1}{\rho(PP^*)}(\log\log \frac{1}{\mu_{\min}}+2\log\frac{4}{\epsilon})$} with $\rho(PP^*)$ lower-bounded as in Equation \eqref{eq:log-Sobolev_binary}.

\begin{proof}(Theorem \ref{thm:convergence_rates_binary}) By Lemma \ref{lem:general_bound_log_Sobolev}, the log-Sobolev constant $\rho(PP^*)$ can be lower-bounded as:

\begin{align*}
        \rho(PP^*)\geq \frac{16\pi^2 \Ap\mu_{\min} p_{\min}^3}{25N^2A^2}\geq \frac{2\pi^2 e^{-4\beta}}{25N^2 A^5},
\end{align*}
where we used that by definition of $P$ in Equation \eqref{eq:transition_prob_payoff_log_linear} and $\mu$ in Equation \eqref{eq:stationary_dist_payoff_log_linear}, $\mu_{\min}$ and $p_{\min}$ can be lower-bounded as follows:

    \begin{align*}
        &\mu_{\min}=\min_{a\in\cAp}\mu(a)\geq \frac{e^{-\beta}}{\Ap}\\
        & P_{a,\Ta}\geq \frac{e^{-\beta}}{N 2A},\quad\forall \Ta\in\cAp(a)\ \Rightarrow\ p_{\min}=\frac{e^{-\beta}}{2A}.
    \end{align*}
Equation \eqref{eq:mixing_time_bound} in Lemma \ref{lem:general_mixing_time} provides the following upper bound on the mixing time:

\begin{align*}
    \tm{P}(\epsilon/4)\leq\frac{1}{\rho(PP^*)}\bigg(\log\log \frac{1}{\mu_*}+2\log\frac{4}{\epsilon}\bigg).
\end{align*}
Plugging the bound on the log-Sobolev constant into this equation we obtain:

\begin{align*}
\tm{P} (\epsilon/4) &\le \frac{25N^2 A^5}{2\pi^2 }e^{4\beta}\bigg(\log\log \frac{1}{\mu_{\min}}+2\log\frac{4}{\epsilon}\bigg)\\
&\le \frac{25N^2 A^5}{2\pi^2 }e^{4\beta}\bigg(\log\log \frac{\Ap}{e^{-\beta}}+2\log\frac{4}{\epsilon}\bigg)\\
&\le \frac{25N^2 A^5}{2\pi^2 }e^{4\beta}\bigg(\log\log \Ap+\log \beta+2\log\frac{4}{\epsilon}\bigg).
\end{align*}
Set $t$ as:

\begin{align}\label{eq:t_payoff}
    t\ge \frac{25N^2 A^5}{2\pi^2 }e^{4\beta}\bigg(\log\log \Ap+\log \beta+2\log\frac{4}{\epsilon}\bigg)
\end{align}
and set $\beta$ as:

\begin{align*}
    \beta\geq \frac{1}{\Delta}\log\left(\left(\Ap-A_*^N\right)\left(\frac{1}{\epsilon A_*^N}-\frac{1}{A_*^N}\right)\right)
\end{align*}

Then, we obtain the following upper bound:

\begin{align*}
        \mathbb{E}[\Phi(a^t)]&=\mathbb{E}_{a\sim\mu^t}[\Phi(a)]\\
        &\geq \mathbb{E}_{a\sim\mu}[\Phi(a)] - 2\|\mu^t-\mu\|_{TV}\max_{a\in\cAp}\Phi(a)\\
        &\geq \max_{a\in\cAp} \Phi(a)-\frac{\epsilon}{2}-\frac{2\epsilon}{4}\\
        &=\max_{a\in\cAp}\Phi(a)-\epsilon,
    \end{align*}
    where the third line follows from Lemma \ref{lem:beta_bound}, the fact that $\|\mu^t-\mu\|_{TV}\le \epsilon/4$ for $t$ set as in Equation \eqref{eq:t_payoff}, and the fact that $\Phi(\cdot)\in[0,1]$. Lemma \ref{lem:beta_bound} is applicable when all players adhere to binary-based log-linear learning rather than log-linear learning since the proof of Lemma \ref{lem:beta_bound} depends only on the stationary distribution $\mu$ of the corresponding learning rule which is the same for log-linear learning and binary log-linear learning. This concludes the proof of Theorem \ref{thm:convergence_rates_binary}.
\end{proof}


\hide{

\section{Robustness of log-linear learning}\label{app:thm3}

In this section, we prove Theorem \ref{thm:General_Convergence} and apply it to the corrupted-utility case to prove Corollary \ref{cor:noise_corruption}.

\subsection{Proof of Theorem \ref{thm:General_Convergence}}

\begin{proof} Consider a learning rule $P$, to prove Theorem \ref{thm:General_Convergence}, we first provide a decomposition that relates the expected value of the potential when the agents follow $P$ defined in Equation \eqref{eq:pertrubed_transitions} to the same quantity when the agents follow $P_\ell$ defined in Equation \eqref{eq:transition_prob_log_linear} instead. 

    We have for all $t, t' \in \bN$ that:
    
    {\small
    \begin{align}
        \bE_{\mu_0 P^t}[\Phi] &= \bE_{\mu_0 P_{\ell}^{t'}}[\Phi] + \bE_{\mu_0 P^t}[\Phi] - \bE_{\mu_0 P_{\ell}^{t'}}[\Phi] \nonumber\\
        &\ge \bE_{\mu_0 P_{\ell}^{t'}}[\Phi] - \sqrt{\Ap} \|P^t - P_{\ell}^{t'}\|_2  \label{ineq:Decomposition}
    \end{align}
    }
    
    where the last line follows because $|\Phi(a)|\le 1$ for all $a\in \cAp$ and because $\|.\|_1 \le \sqrt{\Ap}\|\cdot\|_2$. The rest of the proof is based on controlling $\|P^t - P_{\ell}^{t'}\|_2$ using our mixing time results.
\\
\noindent
    \textbf{Decomposition:} Using Lemma \ref{lem:Lipschitz} we obtain:
    
    {\small
    \begin{align*}
        \|P^t - P_{\ell}^{t'}\|_2 &\le \|P^t - \mu \|_2 + \|P_{\ell}^{t'} - \mu_{\ell}\|_2 + \|\mu - \mu_{\ell}\|_2\\
        &\le \|P^t - \mu \|_2 + \|P_{\ell}^{t'} - \mu_{\ell}\|_2 + L(P_{\ell}) \|P-P_{\ell}\|_2\\
        &\le 2\|P^t - \mu \|_{TV} + \|P_{\ell}^{t'} - \mu_{\ell}\|_2 + L(P_{\ell}) \|P-P_{\ell}\|_2,
    \end{align*}
    }
    
    where $L(P_\ell) = \frac{2 \Ap}{\rho(P_\ell P_\ell^*)} (\log\log \frac{1}{\mu_{\ell,\min}}+\log(8 \Ap))$ follows from Lemma \ref{lem:Lipschitz}. In Theorem \ref{thm:convergence_rates}, we showed that $\mu_{\ell, \min} \geq \frac{e^{-\beta}}{\Ap}$ and $\rho(P_\ell P_\ell^*)\geq \frac{16\pi^2 e^{-4\beta}}{25N^2 A^5}$, therefore 
    
    {\small
    \begin{equation*}
        L(P_\ell) \le \frac{25 N^2 A^{N+5} e^{4 \beta }}{8 \pi^2} (\log\log \Ap e^\beta +\log(8 \Ap)).
    \end{equation*}
    }
    
    Thus, for
    
    {\small
    \begin{align*}
        \begin{cases}
            t &\ge  \tm{P}(\epsilon/(4\sqrt{\Ap}))) \\
            t' &\to \infty\\
            \beta &= \frac{\log\left(\left(\Ap-\Ap(\epsilon/2
)\right)\left(\frac{4}{\epsilon/2 \Ap(\epsilon/2)}-\frac{1}{\Ap(\epsilon/2)}\right)\right)}  {\max\{\epsilon/2
,\Delta\}} 
        \end{cases}
    \end{align*}
    }
    
    we have
    
    {\small
    \begin{align*}
    \begin{cases}
        \|P^t - P_{\ell}^t\|_2 &\le \epsilon / \left(2\sqrt{\Ap}\right) + L(P_{\ell}) \|P-P_{\ell}\|_2 \\
        \bE_{a\sim\mu^0P_{\ell}^{t'}}[\Phi(a)] &\ge \max_{a\in \cAp} \Phi(a) - \epsilon/2, \\
        L(P_\ell) = \mathcal{O}\left( N^2 A^{N+5} e^{\frac{\log(\Ap/\epsilon)}{\max\{\epsilon,\Delta\}}} \left(\log\log \Ap e^{\frac{\log(\Ap/\epsilon)}{\max\{\epsilon,\Delta\}}} + \log(\Ap) \right)\right),
    \end{cases}
    \end{align*}
    }
    
    where the second line follows from Lemma \ref{lem:beta_bound}. Plugging the above inequalities with the bound on $L$ from Lemma \ref{lem:Lipschitz} into the decomposition \eqref{ineq:Decomposition} proves the desired result. We now bound the mixing time $\tm{P}(\epsilon/(4\sqrt{\Ap}))$ of the Markov chain induced by $P$.
    \\
    \noindent
    \textbf{Mixing time bound:} To bound the mixing time of the Markov chain induced by $P$, we use inequality \eqref{eq:mixing_time_bound} and Lemma \ref{lem:general_bound_log_Sobolev}. Assuming a lower bound of $p_{\min}/N$ on the probabilities of all feasible transitions implies a bound on the stationary distribution as we will show next.  
\\
\noindent
    \underline{\textit{Lower bound $(\mu_P)_{\min}$}:} Since $P$ has a positive probability of transitioning from $a\in \cAp$ to any $\Ta \in \cN(a)$, it follows that the corresponding $N$-step transition $P^N$ has a positive probability of transitioning from any $a\in \cAp$ to any $a' \in \cAp$, \ie
    
    {\small
    \begin{equation*}
        \forall a,a' \in \cAp: \quad P^N_{a,a'} \ge N! \:(p_{\min}/N)^N.
    \end{equation*}
    }
    
    Note that the least probable transitions are when $a$ and $a'$ are such that $\forall i \in [N]: \: a_i \neq a'_i$. For all such transitions, the possible paths using $P^N$ are the permutations of $\{1,\ldots,N\}$ (each of the $N$ steps is a new player updating their action). There are $N!$ such permutations and each player $i \in [N]$ can update $a_i$ to $a'_i$ with probability larger than $p_{\min}/N$. 

    Since $P$ is an irreducible and aperiodic transition matrix, the Markov chain induced by $P$ has a unique stationary distribution $\mu_P$. It is known that the Markov chain induced by $P^N$ has the same stationary distribution as the one induced by $P$. Therefore, we have for all $a\in\cAp$:
    
    {\small
    \begin{align*}
        \mu_P (a) &= \sum_{\Ta \in \cAp} P^N_{\Ta,a} \mu_P (\Ta)\\
        &\ge \sum_{\Ta \in \cAp} N! \:(p_{\min}/N)^N \mu_P (\Ta) = N! \:(p_{\min}/N)^N.
    \end{align*}
    }
    
    Therefore, $(\mu_P)_{\min}\ge N! \:(p_{\min}/N)^N$.
\\
\noindent
    \underline{\textit{Deducing the mixing-time bound}:} We can now give an explicit bound on the mixing time of the transition $P$. First, we have by Lemma \ref{lem:general_bound_log_Sobolev}:
    
    {\small
    \begin{align*}
        \rho(PP^*) &\ge \frac{16 \pi^2 \Ap (\mu_P)_{\min} p_{\min}^3}{25 N^2}\\
        &\ge \frac{4\pi^2 \Ap p_{\min}^{N+3} N!}{25N^{N+2}}.
    \end{align*}
    }
    
    Then, using Stirling's formula, we have $N! \ge \sqrt{2 \pi N}\left(\frac{N}{e}\right)^N $, therefore
    
    {\small
    \begin{align*}
        \rho(PP^*) \ge \frac{(2\pi)^{5/2} \Ap p_{\min}^{N+3}}{25 N^{3/2} e^N}
    \end{align*}
    }
    
    Then, using inequality \eqref{eq:mixing_time_bound}, we have:
    
    {\small
    \begin{align*}
        &\tm{P}(\epsilon/(4\sqrt{\Ap}))\\
        &\le \frac{1}{\rho(PP^*)}\left(\log\log \frac{1}{(\mu_P)_{\min}} + 2 \log\frac{4\sqrt{\Ap}}{\epsilon}\right)\\
        &\le \frac{25 N^{3/2} e^N}{(2\pi)^{5/2} \Ap p_{\min}^{N+3}} \left(\log\log\frac{e^N}{p_{\min}^N \sqrt{2\pi N}} + 2 \log\frac{4 \sqrt{\Ap}}{\epsilon}\right).
    \end{align*}
    }
    
    This concludes the proof of Theorem \ref{thm:General_Convergence}.
\end{proof}
}


\hide{
\subsection{Proof of Lemma \ref{lem:Lipschitz}}\label{app:lemma_lipschitz}

\begin{proof}
Denote by $M\in\R^{\Ap\times\Ap}$ the matrix, where each row corresponds to $\mu_1$. For all $t\in \bN$, we have that:

{\small
\begin{align*}
    \mu_1 - \mu_2 &= (P_1^t)^\top (\mu_1 - \mu_2) + (P_1^t - P_2^t)^\top \mu_2\\
    & = (P_1^t - M)^\top (\mu_1 - \mu_2) + M^\top (\mu_1 - \mu_2) \\
    & \quad+ (P_1^t - P_2^t)^\top \mu_2.
\end{align*}
}

This yields:

{\small
\begin{align*}
    \|\mu_1 - \mu_2\|_2 &\le \|(P_1^t - M)^\top (\mu_1 - \mu_2)\|_2 \\
    &\quad+ \|M^\top (\mu_1 - \mu_2)\|_2 + \|(P_1^t - P_2^t)^\top \|_2 \|\mu_2\|_2\\
    &\le \|P_1^t - M\|_2 \|\mu_1 - \mu_2\|_2 \\
    &\quad+ \|M^\top (\mu_1 - \mu_2)\|_2 + \|P_1^t - P_2^t\|_2\\
    &\le 2\sqrt{\Ap}\|P_1^t - M\|_{TV} \|\mu_1 - \mu_2\|_2 \\
    &\quad+ \|M^\top (\mu_1 - \mu_2)\|_2 + \|P_1^t - P_2^t\|_2
\end{align*}
}

where in the last inequality we used the equivalence of $\|\cdot\|_2$ and $\|\cdot\|_1$ and that $\|\cdot\|_1 = 2 \|\cdot\|_{TV}$ by definition of the total variation distance. 
Furthermore:

{\small
\begin{align*}
    M^\top (\mu_1 - \mu_2) &= \bigg(\mu_1(a) \underbrace{\sum_{a'\in \cAp} (\mu_1(a')-\mu_2(a'))}_{=0}\bigg)_{a\in \cAp}\\
    &=0.
\end{align*}
}

Therefore, we obtain that:

{\small
\begin{align}
    \|\mu_1 - \mu_2\|_2 &\le 2\sqrt{\Ap} \|P_1^t - M\|_{TV} \|\mu_1 - \mu_2\|_2 \nonumber\\
    &\quad+ \|P_1^t - P_2^t\|_2. \label{ineq:first_step_lipschitz}
\end{align}
}

Note that for the second term on the right-hand side in the equation above we have:

{\small
\begin{align*}
    P_1^t - P_2^t & = P_1^t + \sum_{l=1}^{t-1}(P_1^{t-l}P_2^l - P_1^{t-l}P_2^l) - P_2^t\\
    & = \sum_{l=1}^{t}(P_1^{t-l+1}P_2^{l-1}  - P_1^{t-l}P_2^l)\\
    & = \sum_{l=1}^{t}(P_1^{t-l} (P_1 - P_2) P_2^{l-1}).
\end{align*}
}

By applying the norm operator, we find that:

{\small
\begin{align*}
    \|P_1^t - P_2^t\|_2 & \le \sum_{l=1}^{t} \|P_1^{t-l}\|_2 \|P_1 - P_2\|_2 \|P_2^{l-1}\|_2\\
    & \le t \Ap \|P_1 - P_2\|_2,
\end{align*}
}

since $\|P\|_2 \le \sqrt{\Ap}$ holds for all transition matrices $P$ over $\cAp$, and in particular for $P_1^{t-l}$ and $P_2^{l-1}$. By plugging the above into inequality \eqref{ineq:first_step_lipschitz} we find:

\begin{align}
    \|\mu_1 - \mu_2\|_2 &\le 2\sqrt{\Ap} \|P_1^t - M\|_{TV} \|\mu_1 - \mu_2\|_2 \nonumber\\
    &\quad+ t \Ap \|P_1 - P_2\|_2. \nonumber
\end{align}
Finally, by choosing $t = t_{\operatorname{mix}}\left(1/\sqrt{16\Ap}\right)$ we find:

\begin{equation*}
    \|\mu_1 - \mu_2\|_2 \le 2 t_{\operatorname{mix}}\left(1/\sqrt{16\Ap}\right) \Ap \|P_1 - P_2\|_2.
\end{equation*}

The proof is then concluded by using the mixing-time bound from inequality \eqref{eq:mixing_time_bound}. 
\end{proof}
}

\section{Robustness of log-linear learning}\label{app:robustness}

In this section, we provide a prood of Lemma \ref{lem:Lipschitz}, Corollaey \ref{cor:noise_corruption}, and Corollary \ref{cor:fixed_share}.

\subsection{Proof of Lemma \ref{lem:Lipschitz}} 
Denote by $M\in\R^{\Ap\times\Ap}$ a matrix where each row corresponds to $\mu_1$. For all $t\in \bN$, we have that:

\begin{align*}
    \mu_1 - \mu_2 &= \langle P_1^t, \mu_1 - \mu_2\rangle + \langle P_1^t - P_2^t, \mu_2\rangle\\
    & = \langle P_1^t - M, \mu_1 - \mu_2\rangle + \langle M,\mu_1 - \mu_2\rangle 
    + \langle P_1^t - P_2^t), \mu_2\rangle.
\end{align*}
This yields:

\begin{align*}
     \|\mu_1 - \mu_2\|_2 &\le \|\langle P_1^t - M,\mu_1 - \mu_2\rangle\|_2 \\
    &\quad+ \|\langle M,\mu_1 - \mu_2\rangle\|_2 + \|(P_1^t - P_2^t)^\top \|_2 \|\mu_2\|_2,
\end{align*}
then,

\begin{align*}
    \|\mu_1 - \mu_2\|_2 &\le \|P_1^t - M\|_2 \|\mu_1 - \mu_2\|_2 \\
    &\quad+ \|\langle M,\mu_1 - \mu_2\rangle\|_2 + \|P_1^t - P_2^t\|_2\\
    &\le 2\sqrt{\Ap}\|P_1^t - M\|_{TV} \|\mu_1 - \mu_2\|_2 \\
    &\quad+ \|\langle M,\mu_1 - \mu_2\rangle\|_2 + \|P_1^t - P_2^t\|_2
\end{align*}
where in the last inequality we used the equivalence of $\|\cdot\|_2$ and $\|\cdot\|_1$ and that by definition of the total variation distance $\|\cdot\|_1 = 2 \|\cdot\|_{TV}$. 
Furthermore: 
\begin{equation*}
    \langle M,\mu_1 - \mu_2\rangle = \big(\mu_1(a) \sum_{a'\in \cAp} (\mu_1(a')-\mu_2(a'))\big)_{a\in \cAp} = 0.
\end{equation*}
Therefore, we obtain that:

\begin{align}
    \|\mu_1 - \mu_2\|_2 &\le 2\sqrt{\Ap} \|P_1^t - M\|_{TV} \|\mu_1 - \mu_2\|_2 + \|P_1^t - P_2^t\|_2. \label{ineq:first_step_lipschitz}
\end{align}
For the second term in the equation above, we have:

\begin{align*}
    P_1^t - P_2^t & = P_1^t + \sum_{l=1}^{t-1}(P_1^{t-l}P_2^l - P_1^{t-l}P_2^l) - P_2^t\\
    & = \sum_{l=1}^{t}(P_1^{t-l} (P_1 - P_2) P_2^{l-1}).
\end{align*}
By applying the norm operator and since $\|P\|_2 \le \sqrt{\Ap}$ holds for all $P$ over $\cAp$ including $P_1^{t-l}$ and $P_2^{l-1}$ we find that:

\begin{align*}
    \|P_1^t - P_2^t\|_2 & \le \sum_{l=1}^{t} \|P_1^{t-l}\|_2 \|P_1 - P_2\|_2 \|P_2^{l-1}\|_2\\
    & \le t \Ap \|P_1 - P_2\|_2.
\end{align*}
Plugging the above in Inequality \eqref{ineq:first_step_lipschitz} we obtain:

\begin{align}
    \|\mu_1 - \mu_2\|_2 &\le 2\sqrt{\Ap} \|P_1^t - M\|_{TV} \|\mu_1 - \mu_2\|_2 \nonumber\\
    &\quad+ t \Ap \|P_1 - P_2\|_2. \nonumber
\end{align}
Finally, by choosing $t = t_{\operatorname{mix}}\left(1/\sqrt{16\Ap}\right)$ we find:

\begin{equation*}
    \|\mu_1 - \mu_2\|_2 \le 2 t_{\operatorname{mix}}\left(1/\sqrt{16\Ap}\right) \Ap \|P_1 - P_2\|_2.
\end{equation*}
We conclude using the mixing-time bound of Inequality \eqref{eq:mixing_time_bound}.

\subsection{Proof of Corollary \ref{cor:noise_corruption}}
\label{app:cor_corruption}

The key idea is to show that the transition matrix of the Markov chain induced by corrupted utilities is close to its corruption-free counterpart.
\\
\begin{proof}
If all players adhere to log-linear learning with corrupted utilities, the induced Markov chain's transition matrix $\hat{P}$ is given, for all $a, \Ta \in \cAp$ by:

\begin{align*}
    \hat{P}_{a,\Ta} &= \frac{1}{N}\frac{e^{\beta \hat{U}_i(\Ta_i,\Ta_{-i})}}{\sum_{a_i'\in\mathcal{A}_i}e^{\beta \hat{U}_i(a_i',\Ta_{-i})}} \mathds{1}_{\Ta \in \cN(a)}, \\
    &= \frac{1}{N}\frac{e^{\beta (U_i(\Ta_i,\Ta_{-i}) + \xi_i(\Ta_i,\Ta_{-i}))}}{\sum_{a_i'\in\mathcal{A}_i}e^{\beta (U_i(a_i',\Ta_{-i}) + \xi_i(a_i',\Ta_{-i}))}} \mathds{1}_{\Ta \in \cN(a)}.
\end{align*}
Since we assumed that the noise is bounded, we can deduce that

\begin{align*}
    P_{a,\Ta} e^{-2 \beta \xi} \le P_{a,\Ta} \le P_{a,\Ta} e^{2 \beta \xi},
\end{align*}
where {\small$P_{a,\Ta} = \frac{1}{N}\frac{e^{\beta U_i(\Ta_i,\Ta_{-i})}}{\sum_{a_i'\in\mathcal{A}_i}e^{\beta U_i(a_i',\Ta_{-i})}} \mathds{1}_{\Ta \in \cN(a)}$} is the transition with the noise-free utility. This entails that

\begin{align*}
    P_{a,\Ta} (e^{-2 \beta \xi} - 1) \le \hat{P}_{a,\Ta} - P_{a,\Ta} \le P_{a,\Ta} (e^{2 \beta \xi} - 1),
\end{align*}
then, since $e^{-2 \beta \xi} - 1 <0$ and $P_{a,\Ta} \le 1/N$ for all $a,\Ta \in \cAp$, we deduce that

\begin{align*}
    (e^{-2 \beta \xi} - 1)/N \le \hat{P}_{a,\Ta} - P_{a,\Ta} \le (e^{2 \beta \xi} - 1)/N,
\end{align*}
and

\begin{align*}
    |\hat{P}_{a,\Ta} - P_{a,\Ta}| \le \frac{1}{N} \max \left\{e^{2 \beta \xi} - 1, 1 - e^{-2 \beta \xi}\right\},
\end{align*}
Finally, since $2\beta\xi \le 1$ and by using that: $1-e^{-x}<x \text{ for } x>0$, and that: $e^{x}-1< \frac{7}{4} x \text{ for } x \in [0,1]$. Then,

\begin{align*}
    |\hat{P}_{a,\Ta} - P_{a,\Ta}| \le \frac{1}{N} \max \left\{\frac{7}{2} \beta \xi, 2\beta \xi \right\} = \frac{7}{2 N} \beta \xi,        
\end{align*}
and finally

\begin{align*}
    \|\hat{P} - P\|_2 &\le \sqrt{\sum_{a,\Ta \in \cAp} \frac{49}{4 N^2} \beta^2 \xi^2}
    = \frac{7 \Ap}{2 N} \beta \xi.
\end{align*}
Also, since $P_{a,\Ta} \ge P_{a,\Ta} e^{-2 \beta \xi}$ and using $P_{a,\Ta} \ge \frac{e^{-\beta}}{NA}$ then we deduce that $P_{a,\Ta} \ge \frac{e^{-\beta(1+2\xi)}}{NA}$.
We conclude the proof with a straightforward application of Theorem \ref{thm:General_Convergence} with $p_{\min} = e^{-\beta(1+2\xi)}/A$ and $\|\hat{P} - P\|_2 \le \frac{7 \Ap}{2 N} \beta \xi$.
\end{proof}

\subsection{Proof of Corollary \ref{cor:fixed_share}}
\label{app:fixed_share}

Similar to Corollary \ref{cor:noise_corruption}, we proceed by showing that the transition matrix of the Markov chain induced by fixed-share log-linear learning is close to that of log-linear learning.
\\
\begin{proof}
If all players adhere to fixed-share log-linear learning, the induced Markov chain's transition matrix $\hat{P}$ is given, for all $a, \Ta \in \cAp$ by:

\begin{align}\label{eq:transition_prob_fixed_share_log_linear}
    \hat{P}_{a,\Ta} = \frac{1}{N} \left(\frac{\xi}{A} + \frac{(1-\xi)e^{\beta U_i(\Ta_i,\Ta_{-i})}}{\sum_{a_i'\in\mathcal{A}}e^{\beta U_i(a_i',\Ta_{-i})}}\right) \mathds{1}_{\Ta \in \cN(a)}.
\end{align}
Then, we have that
\begin{align*}
    \hat{P}_{a,\Ta} \ge \left(\frac{\xi}{NA} + \frac{(1-\xi) e^{-\beta}}{NA}\right)  \mathds{1}_{\Ta \in \cN(a)},
\end{align*}
which entails that $\hat{P}$ satisfies the condition of Theorem \ref{thm:General_Convergence} with $p_{\min} \ge \frac{\xi}{A} + \frac{(1-\xi) e^{-\beta}}{A}$. Additionally, we can show that:

\begin{align*}
    \hat{P}_{a,\Ta} - P_{a,\Ta} &= \frac{1}{N} \left(\frac{\xi}{A} - \frac{\xi e^{\beta U_i(\Ta_i,\Ta_{-i})}}{\sum_{a_i'\in\mathcal{A}}e^{\beta U_i(a_i',\Ta_{-i})}}\right) \mathds{1}_{\Ta \in \cN(a)}\\
    &= \frac{\xi}{N} \left(\frac{1}{A} - \frac{e^{\beta U_i(\Ta_i,\Ta_{-i})}}{\sum_{a_i'\in\mathcal{A}}e^{\beta U_i(a_i',\Ta_{-i})}}\right) \mathds{1}_{\Ta \in \cN(a)},
\end{align*}
where $P$ is the transition matrix of log-linear learning. Therefore,

\begin{align*}
    &\sum_{a,\Ta \in \cAp} \left(\hat{P}_{a,\Ta} - P_{a,\Ta}\right)^2\\
    &= \frac{\xi^2}{N^2} \sum_{a,\Ta \in \cAp} \left(\frac{1}{A^2} - \frac{2e^{\beta U_i(\Ta_i,\Ta_{-i})}}{A \sum_{a_i'\in\mathcal{A}}e^{\beta U_i(a_i',\Ta_{-i})}} + \frac{e^{2 \beta U_i(\Ta_i,\Ta_{-i})}}{ \left(\sum_{a_i'\in\mathcal{A}}e^{\beta U_i(a_i',\Ta_{-i})}\right)^2}\right) \mathds{1}_{\Ta \in \cN(a)}\\
    &\le \frac{\xi^2}{N^2} \sum_{a \in \cAp} \left(\frac{N}{A} - \frac{2N}{A} + N\right) \mathds{1}_{\Ta \in \cN(a)}\\
    &\le N \Ap,
\end{align*}
where the second line follows because from any action profile $a \in \Ap$, there are $NA$ possible transitions ($A$ possible actions times $N$ possible player selections). We also used {\small$\sum_{\Ta \in \cAp}\frac{e^{\beta U_i(\Ta_i,\Ta_{-i})}}{\sum_{a_i'\in\mathcal{A}}e^{\beta U_i(a_i',\Ta_{-i})}} \mathds{1}_{\Ta \in \cN(a)} = 1$} and that {\small$\sum_{\Ta \in \cAp}\frac{e^{\beta U_i(\Ta_i,\Ta_{-i})}}{\left(\sum_{a_i'\in\mathcal{A}}e^{\beta U_i(a_i',\Ta_{-i})}\right)^2} \mathds{1}_{\Ta \in \cN(a)} \le 1$}.

\noindent Finally, since the spectral norm is smaller than the Frobenius norm, then

\begin{align*}
    \|\hat{P} - P\|_2 &\le \sqrt{\sum_{a,\Ta \in \cAp} \left(\hat{P}_{a,\Ta} - P_{a,\Ta}\right)^2}\le \xi \sqrt{\frac{\Ap}{N}}.
\end{align*}
The proof is then concluded by a straightforward application of Theorem \ref{thm:General_Convergence} with $p_{\min} \ge \frac{\xi}{A} + \frac{(1-\xi) e^{-\beta}}{A}$ and $\|\hat{P} - P\|_2 \le \xi \sqrt{\frac{\Ap}{N}}$.

\end{proof}

\end{document}